\documentclass{article}

\usepackage{arxiv}

\usepackage[utf8]{inputenc} 
\usepackage[T1]{fontenc}    
\usepackage[hidelinks]{hyperref}
\usepackage{url}            
\usepackage{booktabs}       
\usepackage{amsfonts}       
\usepackage{nicefrac}       
\usepackage{microtype}      
\usepackage{lipsum}		
\usepackage{graphicx}
\usepackage{natbib}
\usepackage{doi}
\usepackage{amsmath}
\usepackage{newtxtext}
\usepackage{newtxmath}
\usepackage{mathdots}

\newcommand\Rey{\mbox{\textit{Re}}}  
\newcommand\Pran{\mbox{\textit{Pr}}} 
\newcommand\Ma{\mbox{\textit{M}}}    

\title{Identification of cross-frequency interactions in compressible cavity flow using harmonic resolvent analysis}

\author{\href{https://orcid.org/0000-0001-7791-5426}{\includegraphics[scale=0.06]{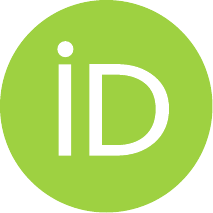}\hspace{1mm}
Md Rashidul Islam}\thanks{Corresponding author: mislam11@syr.edu}, \href{https://orcid.org/0000-0001-7808-672X}{\includegraphics[scale=0.06]{orcid.pdf}\hspace{1mm}Yiyang Sun} \\
	Department of Mechanical and Aerospace Engineering\\
	Syracuse University\\
	Syracuse, NY 13244, USA
}

\renewcommand{\shorttitle}{Identification of cross-frequency interaction using harmonic resolvent analysis}

\hypersetup{
pdftitle={Harmonic resolvent analysis of cavity flow},
pdfsubject={q-bio.NC, q-bio.QM},
pdfauthor={David S.~Hippocampus, Elias D.~Striatum},
pdfkeywords={First keyword, Second keyword, More},
colorlinks = {true},
urlcolor   = {black},
citecolor  = {blue},
}

\begin{document}
\maketitle

\begin{abstract}
The resolvent analysis reveals the worst-case disturbances and the most amplified response in a fluid flow that can develop around a stationary base state. The recent work by \cite{padovan2020analysis} extended the classical resolvent analysis to the harmonic resolvent analysis framework by incorporating the time-varying nature of the base flow. The harmonic resolvent analysis can capture the triadic interactions between perturbations at two different frequencies through a base flow at a particular frequency. The singular values of the harmonic resolvent operator act as a gain between the spatio-temporal forcing and the response provided by the singular vectors. In the current study, we formulate the harmonic resolvent analysis framework for compressible flows based on the linearized Navier-Stokes equation (i.e., operator-based formulation). We validate our approach by applying the technique to the low-mach number flow past an airfoil. We further illustrate the application of this method to compressible cavity flows at Mach numbers of 0.6 and 0.8 with a length-to-depth ratio of $2$. For the cavity flow at Mach number of 0.6, the harmonic resolvent analysis reveals that the nonlinear cross-frequency interactions dominate the amplification of perturbations at frequencies that are harmonics of the leading Rossiter mode in the nonlinear flow. The findings demonstrate a physically consistent representation of an energy transfer from slow-evolving modes toward fast-evolving modes in the flow through cross-frequency interactions. For the cavity flow at Mach number of 0.8, the analysis also sheds light on the nature of cross-frequency interaction in a cavity flow with two coexisting resonances.
\end{abstract}


\section{Introduction}
\label{sec:headings} 
Fluid flows are characterized by a wide range of spatial and temporal structures that result in high dimensional data and pose a significant challenge to their analysis. Fortunately, in many fluid flows, a few structures correlated over space and time, namely the \textit{coherent} structures, drive the underlying physical processes such as mass and energy transport, and sometimes act as a noise source. Linear analysis techniques \citep{taira2017modal,taira2020modal,schmid2007nonmodal} based on data-driven and operator-based methods have been successfully shown to identify those coherent structures and provide additional insight into the nature of instabilities that generate those coherent structures which might develop or already present in a particular flow. The identified structures, also known as the \textit{modes} of the fluid flow, can then be used to build a reduced-complexity model to describe and control the flow dynamics \citep{rowley2017model}.

The operator-based linear tools are traditionally derived from the Navier--Stokes equation (NSE) linearized about a steady solution (or temporal mean). Analyzing the stability of perturbations in viscous parallel shear flows using the eigenspectrum of the linear operator dates back to early 1900 \citep{schmid2002stability}. With the rise in computational power, the analysis can now be performed in a global framework considering two- and three-dimensional base flows \citep{theofilis2011global}. The global stability analysis using a steady solution of the NSE or a time-averaged flow has provided insight into the intrinsic instability mechanisms (e.g., Kelvin--Helmholtz, Tollmien--Schlichting), and the resulting coherent structures in various canonical flows such as jet flow \citep{schmidt2017wavepackets}, cylinder flow \citep{barkley2006linear,noack1994global} and cavity flow \citep{sipp2010dynamics,bres2008three,sun2017biglobal}. However, for many shear flows, the linearized Navier-Stokes (LNS) operator is non-normal, which results in non-orthogonal global modes, and their superposition can give rise to short-time amplification of perturbations even in the absence of unstable global modes \citep{trefethen1993hydrodynamic}. In such cases, the pseudospectral analysis \citep{reddy1993pseudospectra,reddy1993energy} using the resolvent norm of the LNS operator characterizes the nature of stability of the flow more accurately. The first instance of reformulating the linearized Navier-Stokes equation (LNSE) in an input-output analysis framework governed by the resolvent operator as a transfer function for laminar channel flow was done by \cite{jovanovic2005componentwise}. Later, \cite{mckeon2010critical} extended the analysis for turbulent flow where the temporal mean of the flow state is used to build the resolvent operator. \cite{mckeon2010critical} showed that the singular value decomposition (SVD) of the resolvent operator provides a way to identify the dominant amplification mechanism present in the flow by studying the structures of the optimal singular vectors known as the forcing and response modes. The connection between the global stability and resolvent modes and the type of amplification mechanism (modal or non-modal) that both the linear analysis can identify is detailed in the study by \cite{symon2018non}. Henceforth, we will refer to the analysis based on the original resolvent formulation of \cite{mckeon2010critical} as the \textit{classical} one to distinguish from the modified approaches to improve the modeling.

The classical resolvent analysis has been successfully used in the literature for understanding the instability mechanism in different flow configurations, obtaining design guidelines for flow control, and estimating velocity fluctuations in turbulent flow \citep{mckeon2010critical,lesshafft2019resolvent,ribeiro2023triglobal,yeh2019resolvent,liu2021unsteady,amaral2021resolvent}. While the classical resolvent analysis can model the coherent structures well at frequencies that dominant mechanism is present in the nonlinear flow, it fails to do so at frequencies that are generated through nonlinear interactions among the existing frequencies, for example, in the presence of strong oscillatory base flow due to vortex shedding \citep{symon2019tale}. The reason for this is the time-invariant nature of the base flow in the classical resolvent analysis can not resolve the cross-frequency interaction. A potential remedy within the classical framework is to model the nonlinear forcing term at those frequencies by considering triadic interactions between a few resolvent response modes and use that forcing to obtain improved response modes that match the structures obtained from other data-driven analysis \citep{rosenberg2019role,symon2019tale}. Another recent approach by \cite{rigas2021nonlinear} proposed nonlinear input-output analysis using the harmonic balance method to model triadic interactions between some finite number of frequencies. The harmonic balance approach is a popular technique to study the input-output properties of linear time-periodic dynamical systems \citep{wereley1990frequency}. In the context of fluid mechanics, \cite{jovanovic2008h2} applied the approach to study the input-output nature of perturbations in the linearized oscillating channel flow through the $H_2$ norm of the time-periodic system. \cite{franceschini2022identification} used quasi-steady resolvent analysis to study the input-output dynamics of high-frequency perturbation developing over low-frequency periodic base flow. Recent work by \cite{ballouz2024wavelet} considered the wavelet-based resolvent analysis framework to model the time-varying nature of the base flow and the evolution of perturbations localized in time. Analogous extensions of the classical resolvent analysis for spatially periodic base flows are considered in \cite{chavarin2020resolvent}, which was used to obtain design guidelines for optimal placement of riblets to control flow in a turbulent channel. The comprehensive review by \cite{jovanovic2021bypass} is an excellent resource to learn more about such recent extensions of the resolvent-based modeling and control. 

The time-dependent base in a highly unsteady flow allows interactions between the perturbations at different frequencies, and the classical resolvent analysis can not resolve the interactions. To circumvent the limitation, the harmonic resolvent analysis described in \cite{padovan2020analysis} extends the classical resolvent analysis for time-varying base flow. In the harmonic resolvent analysis, the NSE is linearized around a periodically time-varying base flow. Then, a system of coupled equations is obtained by applying Fourier series expansions to the linearized equations together with the harmonic balance approach. More specifically, the perturbations at different frequencies are coupled through the base flow due to its time-varying nature. Using the approach, \cite{padovan2020analysis} reformulated the incompressible NSE in an input-output form between perturbations with a set of coupled frequencies, and their dynamics are governed by the \textit{harmonic resolvent operator} in the frequency domain. The SVD of the harmonic resolvent operator provides insight into the dominant amplification mechanism of perturbations about the time-varying base flow. Although the harmonic resolvent formulation has been provided for the incompressible NSE in \cite{padovan2020analysis}, the need for analyzing perturbation dynamics about high-speed time-varying base flow involving unsteady shock and its interactions with shear-layer/boundary-layer warrants compressibility consideration. Such formalism for the compressible NSE is not available in the literature. While we note that computing harmonic resolvent modes using the time-stepping method \citep{farghadan2024efficient} can be an alternative, formulating the harmonic resolvent framework from the compressible linearized NSE in Fourier space as the matrix-based approach can provide more flexibility in manipulating parametric studies. 

This work derives the harmonic resolvent framework for compressible flows in the frequency domain from the first principle. Using the developed framework, we study the cross-frequency interactions in subsonic open-cavity flows. Flow over an open cavity is a canonical configuration where multiple resonances due to the Rossiter feedback mechanism \citep{rossiter1964wind} drive a self-sustained shear-layer oscillation at multiple frequencies. This problem serves as a nice example of analyzing the perturbation amplification around the time-varying flow using the harmonic resolvent analysis. 

The structure of the paper is as follows. In \S\ref{sec:theory}, we review the classical resolvent analysis and its extension to the harmonic resolvent analysis for a general linear dynamical system. Then, we provide the harmonic resolvent formulation for the compressible NSE and describe the construction of the harmonic resolvent operator. We finish \S\ref{sec:theory} by explaining the connection between the amplification mechanism and the SVD of the harmonic resolvent operator. We validate the formulation in \S\ref{sec:validation} using a problem of flow over an airfoil. We compare the dominant forcing and response modes obtained using our implemented method against the result of \cite{padovan2020analysis}. We then apply the technique in \S\ref{sec:cavity} to study the cross-frequency interactions in cavity flows at Mach numbers of 0.6 and 0.8. Finally, we provide concluding remarks and future considerations in \S\ref{sec:conclusion}. In addition, we give supplemental discussions in the appendices to complement the discussion of the main text.

\section{Theoretical formulation}\label{sec:theory}
In this section, we discuss the mathematical formulation of the harmonic resolvent analysis. We briefly review the classical resolvent analysis formulation followed by its extension to the harmonic resolvent analysis for a general nonlinear dynamical system. Then, we provide the governing equation of perturbations in a time-varying compressible fluid flow and derive the corresponding harmonic resolvent formulation for unsteady base flows in the frequency domain. 

\subsection{Linear-time-varying dynamical system}
We begin with a general dynamical system of the form
\begin{equation}
    \frac{d \boldsymbol q(t)}{dt} = \mathcal N(\boldsymbol q(t)),
    \label{NDS}
\end{equation}
where $\boldsymbol q(t)\in \mathbb R^N$ is the state vector, and $\mathcal N:\mathbb R^N\rightarrow \mathbb R^N$ is a nonlinear function that describes the dynamics of the system. We decompose the state $\boldsymbol q(t)$ into a base state $\boldsymbol Q(t)$ and a perturbation $\boldsymbol q'(t)$ such that $\boldsymbol q(t)= \boldsymbol Q(t) + \boldsymbol q'(t)$. We substitute the decomposition into equation~(\ref{NDS}) and adopt the Taylor series expansion. After neglecting terms that are third-order or higher, we obtain
\begin{equation}
\begin{split}
    \frac{d\boldsymbol Q(t)}{dt} + \frac{d \boldsymbol q'(t)}{dt}
    = \mathcal N(\boldsymbol Q(t)) + \left. \frac{\partial \mathcal N}{\partial \boldsymbol q} \right\vert_{\boldsymbol Q(t)} \boldsymbol q'(t) + \mathcal{O}^2(\boldsymbol q'(t)),
\end{split}
\label{RDecomp}
\end{equation}
which can be cast as
\begin{equation}
    \frac{d \boldsymbol q'(t)}{dt} = \boldsymbol A(t) \boldsymbol q'(t) + \boldsymbol f'(t),
    \label{LDS}
\end{equation}
where $\boldsymbol A(t) = \left. \partial \mathcal N/ \partial \boldsymbol q \right\vert_{\boldsymbol Q(t)}\in \mathbb R^{N\times N}$ is the Jacobian operator, and  
\begin{equation}
    \boldsymbol f'(t) = \mathcal N(\boldsymbol Q(t)) - \frac{d \boldsymbol Q(t)}{dt} + \mathcal{O}^2(\boldsymbol q'(t)),
\end{equation}
which contains the residual terms arising from the equation~(\ref{NDS}) when $\boldsymbol Q(t)$ is not an exact solution and the second order terms that are nonlinear in $\boldsymbol q'(t)$. From the general expression of the linearized perturbation dynamics governed by the equation~(\ref{LDS}), we will derive the classical and harmonic resolvent analysis frameworks in \S\ref{sec:CRA_general} and \S\ref{sec:HRA_general}, respectively, depending on the nature of the base state (i.e., stationary or time-varying). 

\subsection{Classical resolvent analysis framework}\label{sec:CRA_general}
In the classical resolvent analysis, we consider the base state to be a steady solution of the equation~(\ref{NDS}) or a statistically stationary time-averaged mean of the solution. Then equation~(\ref{LDS}) becomes
\begin{equation}
    \frac{d \boldsymbol q'(t)}{dt} = \boldsymbol A\, \boldsymbol q'(t) + \boldsymbol f'(t),
    \label{LTI}
\end{equation}
where $\boldsymbol A\in \mathbb R^{N\times N}$ is the time-invariant Jacobian operator which governs the dynamics of the unsteady perturbations $\boldsymbol q'(t)$ with an unsteady forcing $\boldsymbol f'(t)$. The forcing $\boldsymbol f'(t)$ has different interpretations in the classical resolvent analysis that are worth discussing. If $\boldsymbol Q$ is an exact solution of the equation~(\ref{NDS}) such that $\mathcal N(\boldsymbol Q)=0$, and the perturbations $\boldsymbol q'(t)$ are small enough to discard $\mathcal O^2(\boldsymbol q'(t))$ terms, we can treat $\boldsymbol f'(t)$ solely as an exogenous forcing such as a control input or freestream disturbance \citep{jovanovic2005componentwise}. Such a scenario can arise for fluid flows at low Reynolds numbers. On the other hand, if the perturbations $\boldsymbol q'(t)$ around the base state are large enough (e.g. in turbulent flows), we retain the $\mathcal O^2(\boldsymbol q'(t))$ terms, and $\boldsymbol f'(t)$ has a nonlinear dependence on $\boldsymbol q'(t)$. Then $\boldsymbol f'(t)$ becomes a combination of endogenous forcing originating from the nonlinear interactions of the perturbations and any other presence of external influence in the system \citep{mckeon2010critical}. Following previous studies \citep{towne2018spectral,liu2021unsteady,mckeon2010critical}, we disregard the dependence of $\boldsymbol f'(t)$ on the state perturbation $\boldsymbol q'(t)$ and consider $\boldsymbol f'(t)$ as an unknown forcing term in the subsequent analysis. Substituting the Fourier transform of the perturbation $\boldsymbol q'(t)$ and the forcing $\boldsymbol f'(t)$
\refstepcounter{equation}
$$
  \boldsymbol q'(t) = \int_{-\infty}^{\infty} \hat{\boldsymbol q}'_{\omega} \mathrm{e}^{\mathrm{i} \omega t} \mathrm{d}\omega,\quad
    \boldsymbol f'(t) = \int_{-\infty}^{\infty} \boldsymbol{\hat{f}}'_{\omega} \mathrm{e}^{\mathrm{i} \omega t} \mathrm{d}\omega,
  \eqno{(\theequation{\mathit{a},\mathit{b}})}
$$
into equation~(\ref{LTI}) yields 
\begin{equation}
    \hat{\boldsymbol q}'_{\omega} = \boldsymbol R_{\omega} \boldsymbol{\hat{f}}'_{\omega},
\end{equation}
where $\boldsymbol R_{\omega} \coloneqq (\mathrm{i} \omega \boldsymbol I - \boldsymbol A)^{-1}\in \mathbb C^{N\times N}$ is the classical resolvent operator \citep{jovanovic2005componentwise,mckeon2010critical}, where $\boldsymbol I$ is a identity matrix. The operator $\boldsymbol R_{\omega}$ acts as an open-loop transfer function from the input forcing $\boldsymbol{\hat{f}}'_{\omega}$ to the output response $\hat{\boldsymbol q}'_{\omega}$ at the frequency $\omega$.

\subsection{Harmonic resolvent analysis framework}{\label{sec:HRA_general}}
The harmonic resolvent analysis \citep{padovan2020analysis} extends the classical resolvent analysis for a time-varying system. In the harmonic resolvent analysis, we consider the base state $\boldsymbol Q(t)$ to be time periodic with a fundamental period $T$ and a fundamental frequency $\omega_p =2\pi/T$. Since the Jacobian matrix $\boldsymbol A(t)$ is evaluated about the base state $\boldsymbol Q(t)$, it inherits the time-periodicity with the same period $T$ of the base state. In the present work, we are interested in understanding the dynamics of a $T$-periodic perturbation $\boldsymbol q'(t)$ developing around the periodic base state. The perturbation dynamics need not be exactly $T$-periodic, and the analysis can be generalized for any $nT$-periodic perturbation, for an integer $n$ \citep{padovan2022analysis}. We expand $\boldsymbol A(t)$, $\boldsymbol q'(t)$, and $\boldsymbol f'(t)$ in terms of Fourier series as
\refstepcounter{equation}
$$
    \boldsymbol A(t) = \sum_{k = -\infty}^{k= \infty} \boldsymbol{\hat A}_k \mathrm{e}^{\mathrm{i} k\omega_p t},\quad
    \boldsymbol q'(t) = \sum_{k=-\infty}^{k=\infty} \hat{\boldsymbol q}'_k \mathrm{e}^{\mathrm{i} k\omega_p t},\quad
    \boldsymbol f'(t) = \sum_{k=-\infty}^{k=\infty} \boldsymbol{\hat{f}}'_k \mathrm{e}^{\mathrm{i} k\omega_p t}.
  \eqno{(\theequation{\mathit{a},\mathit{b},\mathit{c}})}
$$
Substituting the Fourier series expansions into equation~(\ref{LDS}) yields
\begin{equation}
    [\boldsymbol T \hat{\boldsymbol q}']_k \coloneqq \mathrm{i} k\omega_p \hat{\boldsymbol q}'_k - \sum_{j=-\infty}^{j=\infty} \boldsymbol{\hat A}_{k-j} \hat{\boldsymbol q}'_j = \boldsymbol{\hat{f}}'_k,~ \forall k,j\in \mathbb Z,
\end{equation}
which represents a system of infinitely coupled equations, where perturbation $\hat{\boldsymbol q}'_k$ at the frequency $k\omega_p$ is coupled with the perturbation $\hat{\boldsymbol q}'_j$ at frequency $j\omega_p$ through the base state at frequency $(k-j)\omega_p$. In a matrix form, we can express the coupled system of equations as
\begin{equation}
    \boldsymbol T \hat{\mathcal Q} = \hat{\mathcal F},
    \label{CoupledEq}
\end{equation}
where $\boldsymbol T$ is an infinite-dimensional Toeplitz matrix of the form 
\begin{equation}
\setlength{\arraycolsep}{2pt}
\renewcommand{\arraystretch}{1.6}
\boldsymbol T = \left[
\begin{array}{ccccccc}
 \ddots &\vdots& \vdots& \vdots& \vdots& \vdots& \iddots \\
  \displaystyle
  \dots& \boldsymbol R_{-2}^{-1} & -\boldsymbol{\hat A}_{-1}& -\boldsymbol{\hat A}_{-2}& -\boldsymbol{\hat A}_{-3}& -\boldsymbol{\hat A}_{-4} & \dots \\
  \displaystyle
  \dots& -\boldsymbol{\hat A}_1 & \boldsymbol R_{-1}^{-1}& -\boldsymbol{\hat A}_{-1}& -\boldsymbol{\hat A}_{-2}& -\boldsymbol{\hat A}_{-3} &\dots  \\
  \displaystyle
  \dots& -\boldsymbol{\hat A}_2& -\boldsymbol{\hat A}_1 & \boldsymbol R_{0}^{-1}& -\boldsymbol{\hat A}_{-1}& -\boldsymbol{\hat A}_{-2} &\dots\\
  \displaystyle
  \dots& -\boldsymbol{\hat A}_3& -\boldsymbol{\hat A}_2& -\boldsymbol{\hat A}_1 & \boldsymbol R_{1}^{-1}& -\boldsymbol{\hat A}_{-1} &\dots\\
  \displaystyle
  \dots& -\boldsymbol{\hat A}_4& -\boldsymbol{\hat A}_3& -\boldsymbol{\hat A}_2& -\boldsymbol{\hat A}_1 &\boldsymbol R_{2}^{-1} &\dots\\
  \displaystyle
  \iddots &\vdots& \vdots& \vdots& \vdots& \vdots& \ddots
\end{array}  \right] ,
\end{equation}
with the infinite-dimensional state perturbation vector and the forcing vector of
\begin{subequations}
\begin{align}
    \setlength{\arraycolsep}{4pt}
        \renewcommand{\arraystretch}{1.5}
        \hat{\mathcal Q} = \left[
        \begin{array}{ccccccc}
         \dots& \hat{\boldsymbol q}'_{-2}& \hat{\boldsymbol q}'_{-1}& \hat{\boldsymbol q}'_{0}& \hat{\boldsymbol q}'_{1}& \hat{\boldsymbol q}'_{2}& \dots
        \end{array}  \right]^T,\\[4pt]
        \setlength{\arraycolsep}{4pt}
        \renewcommand{\arraystretch}{1.5}
        \hat{\mathcal F} = \left[
        \begin{array}{ccccccc}
         \dots& \boldsymbol{\hat{f}}'_{-2}& \boldsymbol{\hat{f}}'_{-1}& \boldsymbol{\hat{f}}'_{0}& \boldsymbol{\hat{f}}'_{1}& \boldsymbol{\hat{f}}'_{2}& \dots
        \end{array}  \right]^T, 
\end{align}
\end{subequations}
respectively. The diagonal of the matrix $\boldsymbol T$ contains the block matrices of the form $\boldsymbol R^{-1}_k \coloneqq (\mathrm{i} k\omega_p \boldsymbol I- \boldsymbol{\hat{A}}_0)\in \mathbb C^{N\times N}$. The off-diagonal blocks of $\boldsymbol T$ are the Fourier components of the Jacobian matrix $\boldsymbol{\hat A}_j\in \mathbb C^{N\times N}$ at the frequency $j\omega_p$ with $j\in \mathbb Z \backslash \{0\}$. Since the matrix $\boldsymbol A(t)$ is real-valued, the Fourier component $\boldsymbol{\hat A}_k$ is the complex conjugate of the component $\boldsymbol{\hat A}_{-k}$ and vice versa. If the base state is steady, then $\boldsymbol{\hat A}_j=0, \forall j\in \mathbb Z \backslash \{0\}$, and the off-diagonal blocks of the matrix $\boldsymbol T$ becomes zero. In the resulting system, the state perturbations at different frequencies are decoupled, and the non-zero diagonal elements of $\boldsymbol T$ are the inverse of the classical resolvent operators at frequencies $k\omega_p$.

Next, we need to define an input-output relation between the forcing $\hat{\mathcal F}$ and state perturbation $\hat{\mathcal Q}$ in the frequency domain using the inverse of the operator $\boldsymbol T$ in equation~(\ref{CoupledEq}). However, if $\boldsymbol Q(t)$ satisfies equation~(\ref{NDS}) exactly, the operator $\boldsymbol T$ is singular and contains a non-zero vector in the right nullspace (see Appendix~\ref{appA}). If, however, the dynamics develop due to external forcing, then the nullspace becomes trivial. The existence of singularity when the null space is non-trivial prevents an inversion of the operator $\boldsymbol T$. Following the work by \citet{padovan2020analysis}, we can restrict the domain and range of the operator $\boldsymbol T$ to remove the singularity. By defining $\hat{\boldsymbol w}$ as a unit norm vector in the right nullspace, we can use the elementary orthogonal projector $\boldsymbol P_{\mathcal{X}} =\boldsymbol I - \hat{\boldsymbol w} \hat{\boldsymbol w}^*$ to project vectors on the subspace $\mathcal X$ that is an orthogonal complement to the right nullspace of $\boldsymbol T$. Here, $(\cdot)^*$ denotes the complex conjugate transpose of a variable. Similarly, we can restrict the range of $\boldsymbol T$ to a subspace $\mathcal{W}$, which is an orthogonal complement of the left nullspace of $\boldsymbol T$ using the elementary projector $\boldsymbol P_{\mathcal{W}} = \boldsymbol I - \hat{\boldsymbol u} \hat{\boldsymbol u}^*$, where $\hat{\boldsymbol u}$ is a unit norm vector in the left nullspace of $\boldsymbol T$. The computation of the unit norm vectors and the associated projection operators in practical applications are detailed in the study by \cite{padovan2020analysis}. The restricted operator $\boldsymbol T_w: \mathcal{X}\rightarrow \mathcal{W}$ is invertible and the input-output relation from equation~(\ref{CoupledEq}) is obtained as
\begin{equation}
    \hat{\mathcal Q} = \boldsymbol H \hat{\mathcal F}.
\end{equation}
where $\boldsymbol H \coloneqq \boldsymbol T_w^{-1}$ is the harmonic resolvent operator. The operator $\boldsymbol H$ maps the Fourier coefficients of the $T$-periodic forcing $\hat{\mathcal F}$ to the Fourier coefficients of the output state perturbation $\hat{\mathcal Q}$ of the same period $T$. 

\subsection{Harmonic resolvent formulation for compressible flow}
\subsubsection{Navier-Stokes equation}
In this section, we derive the harmonic resolvent formulation for the fluid flow governing equations. We consider the compressible Navier--Stokes equation in the conservative formulation. The Cartesian coordinate system $x_i~(i=1, 2, 3)$, time $t$, density $\rho$, three components of velocity $u_i$, pressure $p$, temperature $T$, and total energy $E$ are non-dimensionalized as
\begin{equation}
    x_i = \frac{\tilde x_i}{L}, t = \frac{\tilde t}{L/ \tilde u_{\infty}}, \rho = \frac{\tilde\rho}{\tilde \rho_{\infty}}, u_i = \frac{\tilde u_i}{\tilde u_{\infty}}, p = \frac{\tilde p}{\tilde \rho_{\infty} \tilde u_{\infty}^2}, T =\frac{\tilde T}{\tilde T_{\infty}}, E = \frac{\tilde E}{\tilde \rho_{\infty} \tilde u_{\infty}^2},
    \label{dimensional}
\end{equation}
where the variables denoted with the symbol $\tilde{(\cdot)}$ are the dimensional quantities, the variables with the subscript $\infty$ denote free-stream values, and $L$ is a dimensional reference length. We introduce three dimensionless numbers namely the Reynolds number $\Rey$, the Prandtl number $\Pran$, and the Mach number $\Ma$
\refstepcounter{equation}
$$
  \Rey = \frac{\tilde \rho_{\infty} \tilde u_{\infty} L}{\tilde \mu_{\infty}},\quad
  \Pran = \frac{\tilde \mu_{\infty} \tilde c_p}{\tilde \kappa},\quad
  \Ma = \frac{\tilde u_{\infty}}{\tilde a_{\infty}}, 
  \eqno{(\theequation{\mathit{a},\mathit{b},\mathit{c}})}
$$
where $\tilde \mu_{\infty}$ is the freestream dynamic viscosity, $\tilde a_{\infty}$ is the speed of sound in the freestream, $\tilde c_p$ is the specific heat at constant pressure, and $\tilde \kappa$ is the thermal conductivity of the fluid. Then we can compactly write the Navier-Stokes equation in a non-dimensional form as 
\begin{equation}
    \frac{\partial \boldsymbol q}{\partial t} + \frac{\partial \boldsymbol F_j^e}{\partial x_j} + \frac{\partial \boldsymbol F_j^v}{\partial x_j} = 0 ,
    \label{NSE}
\end{equation}
where $\boldsymbol q = [\rho, m_i, \rho E]^T\in \mathbb{R}^5$ is the vector of the conservative state variables with $m_i\coloneqq \rho u_i$ being the three components of the momentum, $\boldsymbol F_j^e$ and $\boldsymbol F_j^v$ represents the Euler flux and viscous flux vector, respectively. For a thermally and calorically perfect gas, total energy $E$ is given by 
\begin{equation}
    E = \frac{p}{\rho(\gamma-1)} + \frac{1}{2} u_k u_k,
\end{equation}
where $\gamma$ is the ratio of specific heat. The Euler flux $\boldsymbol F_j^e$ and the viscous flux $\boldsymbol F_j^v$ \citep{bugeat20193d} are given by 
\refstepcounter{equation}
$$
  \setlength{\arraycolsep}{0pt}
    \renewcommand{\arraystretch}{1.4}
    \boldsymbol F_j^e = \left[
    \begin{array}{c}
       m_j\\
       \displaystyle
       m_i u_j + p \delta_{ij}\\
       \displaystyle
       (\rho E+p)u_j
    \end{array}  \right] ,\quad
    \setlength{\arraycolsep}{0pt}
    \renewcommand{\arraystretch}{1.4}
    \boldsymbol F_j^v = \left[
    \begin{array}{c}
       0\\
       \displaystyle
       -\frac{1}{\Rey}\, \tau_{ij}\\
       \displaystyle
       -\frac{1}{\Rey}\, u_i \tau_{ij} - \frac{\mu}{(\gamma-1)\Rey\, \Pran\, \Ma^2}\, \frac{\partial T}{\partial x_j}
    \end{array}  \right] ,
  \eqno{(\theequation{\mathit{a},\mathit{b}})}
  \label{flux}
$$
where 
\begin{equation}
    \tau_{ij} = \mu \left(\frac{\partial u_i}{\partial x_j} + \frac{\partial u_j}{\partial x_i} - \frac{2}{3} \frac{\partial u_k}{\partial x_k} \delta_{ij}\right)
\end{equation}
is the viscous stress tensor for a Newtonian fluid. Note that the coefficients of the viscous stress term and the temperature gradient term in equation~(\ref{flux}$\mathit{b}$) depend on the reference variables used to non-dimensionalize the governing equations. The dynamic viscosity $\mu$ is a function of the temperature, which is calculated using a power law as $\mu(T) = (T/T_{\infty})^{0.76}$ \citep{garnier2009large}. We use the equation of state to relate the pressure, density, and temperature as
\begin{equation}
    p = \frac{1}{\gamma\, \Ma^2} \rho T.
\end{equation}
We take the value of the Prandtl number  $\Pran=0.7$ and the specific heat ratio $\gamma=1.4$, which are the standard values for air.

\subsubsection{Linearized Navier-Stokes equation}
We decompose the state variables $\boldsymbol q(t)$ into a periodic base state $\overline{\boldsymbol q}(t)$ and an unsteady perturbation $\boldsymbol q'(t)$ as $\boldsymbol q(t) = \overline{\boldsymbol q}(t) + \boldsymbol q'(t)$. Substituting the decomposition into the equation~(\ref{NSE}) and linearizing around the base state, we obtain 
\begin{equation}
    \frac{\partial \boldsymbol q'(t)}{dt} + \frac{\partial}{\partial x_j} \mathcal F_j^e(\overline{\boldsymbol q}(t),\boldsymbol q'(t)) + \frac{\partial}{\partial x_j}  \mathcal F_j^v(\overline{\boldsymbol q}(t),\boldsymbol q'(t)) = \boldsymbol f'(t),
    \label{LNSE}
\end{equation}
where $\boldsymbol q'(t) =[\rho',m'_i,\rho E']$ is the vector of conservative variable perturbations and $\boldsymbol f'(t)$ contains the terms which are nonlinear in $\boldsymbol q'(t)$ and the residual terms if $\overline{\boldsymbol q}(t)$ is not an exact solution of equation~(\ref{NSE}) as follows 
\begin{equation}
    \boldsymbol f'(t) = -\frac{\partial \overline{\boldsymbol q}(t)}{dt} - \frac{\partial}{\partial x_j} \mathcal F_j^e(\overline{\boldsymbol q}(t)) - \frac{\partial}{\partial x_j} \mathcal F_j^v(\overline{\boldsymbol q}(t)) -\mathcal{O}^2(\boldsymbol q'(t)).
\end{equation}
The linearized Euler flux $\mathcal F_j^e(\overline{\boldsymbol q}(t),\boldsymbol q'(t))$ read
\begin{equation}
\setlength{\arraycolsep}{0pt}
\renewcommand{\arraystretch}{2.0}
\mathcal F_j^e(\overline{\boldsymbol q}(t),\boldsymbol q'(t)) = \left[
\begin{array}{c}
    m'_j\\
   \displaystyle
   \frac{\overline m_i}{\overline \rho} m'_j + \frac{\overline m_j}{\overline \rho} m'_i - \frac{\overline m_i \overline m_j}{\overline{\rho}^2} \rho' + p'\delta_{ij}\\
   \displaystyle
   \left(\gamma \overline{\rho E} -\frac{\gamma-1}{2} \frac{\overline m_k \overline m_k}{\overline \rho} \right) u'_j + \frac{\overline m_j}{\overline \rho} (\rho E)' + \frac{\overline m_j}{\overline \rho} p' 
\end{array}  \right] ,
\end{equation}
and $\mathcal F_j^v(\overline{\boldsymbol q}(t),\boldsymbol q'(t))$ is the linearized viscous flux vector with 
\begin{equation}
\setlength{\arraycolsep}{0pt}
\renewcommand{\arraystretch}{2.0}
\mathcal F_j^v(\overline{\boldsymbol q}(t),\boldsymbol q'(t)) = \left[
\begin{array}{c}
    0\\
   \displaystyle
   - \frac{1}{Re} \tau'_{ij}\\
   \displaystyle
   -\frac{1}{\Rey}\, u'_i \overline{\tau}_{ij} -\frac{1}{\Rey}\, \overline u_i \tau'_{ij} - \frac{\overline \mu}{(\gamma-1)\Rey\, \Pran\, \Ma^2}\, \frac{\partial T'}{\partial x_j}
\end{array}  \right] ,
\end{equation}
where
\refstepcounter{equation}
$$
     \overline \tau_{ij} = \overline \mu \left(\frac{\partial \overline u_i}{\partial x_j} + \frac{\partial \overline u_j}{\partial x_i} - \frac{2}{3} \frac{\partial \overline u_k}{\partial x_k} \delta_{ij}\right),\quad
     \tau'_{ij} = \overline \mu \left(\frac{\partial u'_i}{\partial x_j} + \frac{\partial u'_j}{\partial x_i} - \frac{2}{3} \frac{\partial u'_k}{\partial x_k} \delta_{ij}\right),
  \eqno{(\theequation{\mathit{a},\mathit{b}})}
$$
and $\overline u_i \coloneqq \overline m_i/\overline \rho$ denotes the velocity components of the base state. We neglect the terms with viscosity perturbation $\mu'$ by assuming its negligible variation with temperature. The perturbations of primitive variable velocity $u_i'$, pressure $p'$, and temperature $T'$ is calculated respectively as
\begin{subequations}
    \begin{align}
      u_i' &= \frac{1}{\overline \rho} m_i' - \frac{\overline m_i}{\overline{\rho}^2} \rho',\\[2pt]
      p' &= (\gamma-1)\left[(\rho E)' - \frac{\overline m_k}{\overline \rho} m_k' +\frac{1}{2} \frac{\overline m_k \overline m_k}{ \overline{\rho}^2} \rho'\right],\\[2pt]
      T' &= \gamma (\gamma-1) \Ma^2 \left[\frac{(\rho E)'}{\overline \rho} - \frac{\overline{\rho E}}{\overline{\rho}^2}\rho' - \frac{\overline m_k}{\overline{\rho}^2} m'_k + \frac{\overline m_k \overline m_k}{\overline{\rho}^3} \rho'\right].
    \end{align}
\end{subequations}
After substituting all the expressions into equation~(\ref{LNSE}), we obtain the governing equation of unsteady perturbations developing over a time-varying compressible fluid flow in the time domain.

\subsubsection{Construction of operator \textit{T}}
To facilitate the conversion of the linearized Navier--Stokes equation from the time domain to the frequency domain, we rewrite equation~(\ref{LNSE}) as
\begin{equation}
     \frac{d \boldsymbol q'(t)}{dt} + \boldsymbol L \boldsymbol q'(t) + \frac{\partial}{\partial x_j} \left[\mathcal G_j^e(\overline{\boldsymbol q}(t),\boldsymbol q'(t)) + \mathcal F_j^v(\overline{\boldsymbol q}(t),\boldsymbol q'(t))\right] = \boldsymbol f'(t),
     \label{LNSE2}
\end{equation}
where we have grouped the terms of the linearized Euler flux that contain the product between base state and perturbation state variables in $ \mathcal G_j^e(\overline{\boldsymbol q}(t),\boldsymbol q'(t))$ and the rest of the terms containing only the perturbation state variables in $\boldsymbol L\boldsymbol{q}'(t)$ (expressions are given in Appendix~\ref{appB}). Then, we expand the periodic base state and perturbation using the Fourier series as
\refstepcounter{equation}
$$
    \overline{\boldsymbol q}(t) = \sum_{k \in \Omega} \hat{\overline{\boldsymbol q}}_k \mathrm{e}^{\mathrm{i} k\omega_p t},\quad
    \boldsymbol q'(t) = \sum_{k\in \tilde{\Omega}} \hat{\boldsymbol q}'_k \mathrm{e}^{\mathrm{i} k\omega_p t},
  \eqno{(\theequation{\mathit{a},\mathit{b}})}
  \label{FS}
$$
where both $\Omega,\tilde{\Omega} \subseteq \{k\omega_p\}\,\forall k\in\mathbb{Z}$, are sets of integer multiples of the fundamental frequency $\omega_p = 2\pi/T$ with $T$ being the fundamental period of the base flow. While one can consider an infinite number of frequencies for the base state and the perturbation, in practical computation, we truncate the number of frequencies in the sets $\Omega$ and $\tilde{\Omega}$ to a finite extent. Usually, $\Omega =\{-m,\dots,-1,0,1,\dots,m\}\omega_p$ contains a small number of frequencies associated with the dominant frequency $\omega_p$ and its harmonics present in the base flow. These frequencies approximate the dominant dynamics of the large-scale coherent structures in the fluid flows. Then we seek to study the dynamics of perturbations with frequencies in the set of $\tilde{\Omega}=\{-n,\dots,-1,0,1,\dots,n\}\omega_p$, where $n$ is chosen by the maximum frequency of perturbation that one wishes to resolve and $n\geq m$. Substituting the Fourier expansions into equation~(\ref{LNSE2}) we obtain the following system of a finite number of coupled equations
\begin{equation}
    \left[\boldsymbol T\hat{\boldsymbol q}'\right]_k = \boldsymbol{\hat{f}}'_k,
\end{equation}
with
\begin{equation}
    \left[\boldsymbol T\hat{\boldsymbol q}'\right]_k = \mathrm{i}k\omega_p \hat{\boldsymbol q}'_k + \boldsymbol L \hat{\boldsymbol q}'_k + \frac{\partial}{\partial x_j} \sum_{\substack{l\in \tilde{\Omega}\\(k-l)\in\Omega}} \left[\hat{\mathcal G}_j^e(\hat{\overline{\boldsymbol q}}_{k-l},\hat{\boldsymbol q'}_l) + \hat{\mathcal F}_j^v(\hat{\overline{\boldsymbol q}}_{k-l},\hat{\boldsymbol q'}_l)\right],
    \label{coupledeq2}
\end{equation}
where the number of equations is linked to the number of perturbation frequencies in the set $\tilde{\Omega}$. To assemble the matrix $\boldsymbol T$, we need to find the expressions for the equations corresponding to the frequencies in $\tilde \Omega$ one at a time. For simplicity, we will consider a set of base flow frequencies $\Omega = \{-1,0,1\}\omega_p$ along with a set of perturbation frequencies $\tilde{\Omega} = \{-2,-1,0,1,2\}\omega_p$ to demonstrate the construction of the operator $\boldsymbol T$. The equation corresponding to the perturbation frequency $-2\omega_p$ can be obtained as
\begin{eqnarray}
  \left[\boldsymbol T\hat{\boldsymbol q}'\right]_{-2} = -\mathrm{i}2\omega_p \hat{\boldsymbol q}'_{-2} + \boldsymbol L \hat{\boldsymbol q}'_{-2} + \frac{\partial}{\partial x_j} \left[\hat{\mathcal G}_j^e(\hat{\overline{\boldsymbol q}}_{0},\hat{\boldsymbol q'}_{-2}) + \hat{\mathcal F}_j^v(\hat{\overline{\boldsymbol q}}_{0},\hat{\boldsymbol q'}_{-2})\right]\nonumber\\
  + \frac{\partial}{\partial x_j} \left[\hat{\mathcal G}_j^e(\hat{\overline{\boldsymbol q}}_{-1},\hat{\boldsymbol q'}_{-1}) + \hat{\mathcal F}_j^v(\hat{\overline{\boldsymbol q}}_{-1},\hat{\boldsymbol q'}_{-1})\right],
  \label{freq_coupling}
\end{eqnarray}
where we have neglected the terms containing $\hat{\overline{\boldsymbol q}}_{k-l}$ with $(k-l)\not \in \Omega$ in the expansion of the sum. For brevity, we show how to perform the Fourier expansion of the terms in $\hat{\mathcal G}_j^e(\hat{\overline{\boldsymbol q}}_{k-l},\hat{\boldsymbol q'}_{l})$ for the linearized compressible NSE in Appendix~\ref{appB}. Similarly, for other frequencies $k\omega_p$ in the set $\tilde \Omega$ we can obtain the expression for $\left[ \boldsymbol T\hat{\boldsymbol q}'\right]_k$ using equation~(\ref{coupledeq2}). Then the system of equations in a matrix form is 
\begin{equation}
\setlength{\arraycolsep}{4pt}
\renewcommand{\arraystretch}{1.6}
\boldsymbol T = \left[
\begin{array}{ccccc}
  \boldsymbol R_{-2}^{-1} & \boldsymbol{\hat G}_{-1}& \boldsymbol 0& \boldsymbol 0& \boldsymbol0  \\
  \displaystyle
   \boldsymbol{\hat G}_1 & \boldsymbol R_{-1}^{-1}& \boldsymbol{\hat G}_{-1}& \boldsymbol 0& \boldsymbol0  \\
  \displaystyle
  \boldsymbol 0 & \boldsymbol{\hat G}_1 & \boldsymbol R_{0}^{-1}& \boldsymbol{\hat G}_{-1}& \boldsymbol 0\\
  \displaystyle
   \boldsymbol 0 & \boldsymbol 0 & \boldsymbol{\hat G}_1 & \boldsymbol R_{1}^{-1}& \boldsymbol{\hat G}_{-1}\\
  \displaystyle
   \boldsymbol 0 & \boldsymbol 0 & \boldsymbol 0 & \boldsymbol{\hat G}_1 & \boldsymbol R_{2}^{-1}
\end{array}  \right] ,
\end{equation}
where
\begin{subequations}
    \begin{align}
        \hat{\boldsymbol G}_k \hat{\boldsymbol q}' &\coloneqq \frac{\partial}{\partial x_j} \left[\hat{\mathcal G}_j^e(\hat{\overline{\boldsymbol q}}_{k},\hat{\boldsymbol q'}) + \hat{\mathcal F}_j^v(\hat{\overline{\boldsymbol q}}_{k},\hat{\boldsymbol q'})\right]\label{eqn:G}\\[4pt]
        \boldsymbol R^{-1}_k &\coloneqq (\mathrm{i}k\omega_p\boldsymbol I + \boldsymbol L + \hat{\boldsymbol G}_0).
    \end{align}
\end{subequations}
The operator $\boldsymbol T$ has dimension $\mathbb{C}^{5N_f\times5N_f}$, where $N_f$ is the number of frequencies in the set $\tilde \Omega$. The number of non-zero blocks in each row of the operator $\boldsymbol T$ depends on the number of base flow frequencies in the set $\Omega$. To numerically solve the system of equations, we discretize equation~(\ref{coupledeq2}) using a finite volume scheme in the present work and obtain the discrete matrices $\boldsymbol L\in \mathbb{R}^{5N_g\times 5N_g}$ and $\hat{\boldsymbol G}_k \in\mathbb{C}^{5N_g\times 5N_g}$ that constitutes the blocks of the operator $\boldsymbol T$, with $N_g$ being the number of discrete grid points. We note that only the matrices $\hat{\boldsymbol G}_k$ need to be constructed, and $\hat{\boldsymbol G}_{-k}$ can be obtained by taking the complex conjugate of $\hat{\boldsymbol G}_{k}$. After assembling the discrete operator $\boldsymbol T$, we can remove its singularity, if necessary, using the method outlined in \S~\ref{sec:HRA_general}. Then the input-output perturbation dynamics of a time-periodic fluid flow in the frequency space are represented by 
\begin{equation}
    \hat{\mathcal{Q}} = \boldsymbol H \hat{\mathcal{F}}
    \label{HR_compressible}
\end{equation}
where, $\boldsymbol H \in \mathbb C^{{(5N_g)N_f\times(5N_g)N_f}}$ is the discrete harmonic resolvent operator, $\hat{\mathcal Q}\in \mathbb{C}^{(5N_g)N_f\times 1}$ contains the collection of Fourier coefficients of the discrete state variable perturbations, and $\hat{\mathcal F}\in \mathbb{C}^{(5N_g)N_f\times 1}$ represents the Fourier coefficients of the discrete forcing variables. 

\subsection{Modal decomposition of the harmonic resolvent operator}\label{sec:modal}
We seek to obtain a reduced-order representation of the input-output dynamics using the modal decomposition of the harmonic resolvent operator. In particular, we need to identify the most amplified output perturbation and the corresponding input perturbation characterized by a gain describing the amplification level. We can define the gain as a ratio of the output to input perturbation energy. To measure the perturbation energy, we introduce the discrete inner products $\langle \boldsymbol{\hat q}',\boldsymbol{\hat q}' \rangle_{q} = \boldsymbol{\hat q}^{'*} \boldsymbol W_{q}\, \boldsymbol{\hat q}'$ and $\langle \boldsymbol{\hat f}',\boldsymbol{\hat f}'\rangle_{f} = \boldsymbol{\hat f}^{'*} \boldsymbol W_{f} \boldsymbol{\hat f}'$ in the output and input space, respectively. Then using the inner product, we can define a norm to measure the perturbation energy in both spaces. The positive definite weight matrices $\boldsymbol W_{q}$ and $\boldsymbol W_{f}$ depend on the choice of energy norm one wishes to optimize. For compressible flows, a widely used measure of the perturbation energy is given by Chu's norm \citep{chu1965energy,hanifi1996transient}, which in the non-dimensional form is
\begin{equation}
    ||\boldsymbol{\hat q}'_p||_E^2 = \boldsymbol{\hat q}_p^{' *} \boldsymbol W_{C}\boldsymbol{\hat q}'_p  = E_\text{Chu} = \int_{\Omega} \boldsymbol{\hat q}_p^{' *} \text{diag}\left(\frac{\overline T_0}{\gamma \Ma^2 \overline \rho_0},\overline \rho_0,\frac{\overline \rho_0}{\gamma(\gamma-1)\Ma^2 \overline T_0}\right) \boldsymbol{\hat q}'_p \mathrm{d}\Omega\, , 
\end{equation}
where $\boldsymbol q_p = [\rho',u'_i,T']^T$ is the vector of primitive variable perturbation and $\hat{\boldsymbol q}'_p$ are the Fourier coefficients. The variables denoted with $(\cdot)_0$ represent the time-averaged quantity. Since the harmonic resolvent formulation is derived using the conservative state variable, we need to modify equation~(\ref{HR_compressible}) to accommodate the primitive variable perturbations before applying Chu's norm. The transformed equation in the primitive state variable reads
\begin{equation}
    \hat{\mathcal Q}_p = \underbrace{\boldsymbol M^{-1} \boldsymbol H \boldsymbol M}_{\boldsymbol H_p} \hat{\mathcal F}_p,
\end{equation}
where $\hat{\mathcal Q}_p$ is the vector of Fourier coefficients of the output perturbation $\boldsymbol q'_p(t)$, $\hat{\mathcal F}_p$ contains the Fourier coefficients of $\boldsymbol f'_p(t)$. The operator $\boldsymbol H_p$ governs the input-output dynamics of the primitive state variable perturbations and the details of the matrix $\boldsymbol M$ are given in Appendix~\ref{appC}. We seek to maximize the gain
\begin{equation}
    \Gamma^2 = \max_{\hat{\mathcal F}_p} \frac{||\hat{\mathcal Q}_p||_E^2}{||\hat{\mathcal F}_p||_E^2} = \max_{\hat{\mathcal F}_p} \frac{\langle \hat{\mathcal Q}_p,\hat{\mathcal Q}_p \rangle_{q}}{\langle \hat{\mathcal F}_p,\hat{\mathcal F}_p \rangle_{f}} = \max_{\hat{\mathcal F}_p} \frac{\hat{\mathcal Q}^*_p \boldsymbol W_{C}\, \hat{\mathcal Q}_p}{\hat{\mathcal F}_p^* \boldsymbol W_{C}\, \hat{\mathcal F}_p},
    \label{gain}
\end{equation}
where we use the same measure of perturbation energy using Chu's norm (i.e., $\boldsymbol W_c$ contain the Chu's norm weights along with the discrete integration weights) in both input and output space leading to $\boldsymbol W_q = \boldsymbol W_f = \boldsymbol W_c$. It is not necessary to use the same energy norm in both spaces. Since $\boldsymbol W_c$ is a symmetric positive definite matrix, we can perform a Cholesky factorization as $\boldsymbol W_c = \boldsymbol N^* \boldsymbol N$. Using the factorization in equation~(\ref{gain}), we can transform Chu's energy norm to a discrete $L_2$ norm as
\begin{eqnarray}
  \Gamma^2 = \max_{\hat{\mathcal F}_p} \frac{\hat{\mathcal F}_p^* \boldsymbol{H}_p^* \boldsymbol N^* \boldsymbol N \boldsymbol H_p \hat{\mathcal F}_p}{\hat{\mathcal F}_p^* {\boldsymbol N}^* \boldsymbol N \hat{\mathcal F}_p} &=& \max_{\hat{\mathcal U}_p} \frac{\hat{\mathcal U}_p^* \boldsymbol N^{-1,*} \boldsymbol{H}_p^* {\boldsymbol N}^* \boldsymbol N \boldsymbol H_p \boldsymbol N^{-1} \hat{\mathcal U}_p}{\hat{\mathcal U}_p^* \hat{\mathcal U}_p }\quad \text{with}\quad \hat{\mathcal{U}}_p = \boldsymbol N \hat{\mathcal{F}}_p \nonumber\\
  &=& \max_{\hat{\mathcal U}_p} \frac{||\boldsymbol N \boldsymbol H_p \boldsymbol N^{-1} \hat{\mathcal{U}}_p||_2^2}{||\hat{\mathcal{U}}_p||_2^2} \nonumber\\
  &=& ||\boldsymbol N \boldsymbol H_p \boldsymbol N^{-1}||_2^2 = \sigma_1^2.
  \label{gainSV}
\end{eqnarray}
So the solution to the optimization problem can be obtained by the SVD of the weighted harmonic resolvent operator 
\begin{equation}
    \boldsymbol N \boldsymbol H_p \boldsymbol N^{-1} = \tilde{\boldsymbol U} \boldsymbol \Sigma \tilde{\boldsymbol V}^*,
\end{equation}
where $\boldsymbol \Sigma = \text{diag}(\sigma_1,\sigma_2,\dots)$ contains the ranked singular values of the operator $\boldsymbol N \boldsymbol H_p \boldsymbol N^{-1}$ in descending order and the maximum energy gain $\Gamma^2$ is given by the leading singular value $\sigma_1^2$. The response modes are the columns of the matrix $\boldsymbol U = \boldsymbol N^{-1} \tilde{\boldsymbol U}$ and the forcing modes are the columns of the matrix $\boldsymbol V = \boldsymbol N^{-1} \tilde{\boldsymbol V}$. The optimal forcing and response modes are given by the first column of the matrix $\boldsymbol V$ and $\boldsymbol U$, respectively. The forcing and response modes are orthonormal in their respective inner products, that is, $\boldsymbol V^* \boldsymbol W_c \boldsymbol V = \boldsymbol I$ and $\boldsymbol U^* \boldsymbol W_c \boldsymbol U = \boldsymbol I$. Next, we recover the complete decomposition of the operator as 
\begin{equation}
    \boldsymbol H_p = \boldsymbol U \boldsymbol{\Sigma} \boldsymbol V^* \boldsymbol W_c
\end{equation}
allowing the output response to be expanded as
\begin{equation}
    \hat{\mathcal{Q}}_p = \sum_{k} \boldsymbol U_k \sigma_k \lambda_k \quad \text{with} \quad \lambda_k = \boldsymbol V_k^* \boldsymbol W_c \hat{\mathcal F}_p.
\end{equation}
If the operator $\boldsymbol H_p$ is low-rank and $\sigma_1 \gg \sigma_2$, then we can use a rank-1 approximation to get a reduced-order representation of the dynamics
\begin{equation}
    \hat{\mathcal{Q}}_p = \boldsymbol U_1 \sigma_1 \lambda_1 \quad \text{with} \quad \lambda_1 = \boldsymbol V_1^* \boldsymbol W_c \hat{\mathcal F}_p.
\end{equation}
If the projection $\lambda_1$ of the input on the forcing mode $\boldsymbol V_1$ is maximal, the output response will have structures similar to the response mode $\boldsymbol U_1$ scaled by the singular value $\sigma_1$. In other words, if we want to excite the optimal response, our input to the system needs to be aligned as closely as possible to the optimal forcing mode. The reduced-order representation has profound significance in understanding the physics of the time-periodic fluid flows and developing inputs for flow control.

We have seen till now that the singular value decomposition of the weighted harmonic resolvent operator sheds light on the global energy amplification mechanism in the time-periodic flow. However, since the time-periodic base flow admits cross-frequency interaction between perturbations, it is possible to study the energy amplification between a pair of input and output at different frequencies \citep{padovan2020analysis}. In particular, we want to maximize the gain between the input energy perturbation $\hat{\boldsymbol f}'_j$ at the frequency $j\omega_p\in\tilde \Omega$ to the output response $\hat{\boldsymbol q}'_{k}$ at the frequency $k\omega_p\in\tilde \Omega$. As before shown in equation~(\ref{gainSV}), the solution to the optimization leads to the singular value decomposition of the weighted operator $\boldsymbol H^w_{j,k}\in \mathbb C^{5N_g\times 5N_g}$, which is the corresponding block of the operator $\boldsymbol N \boldsymbol H_p \boldsymbol N^{-1}$ that couples the input $\hat{\boldsymbol f}'_j$ to the output $\hat{\boldsymbol q}'_k$ as  
\begin{equation}
    \hat{\boldsymbol q'}_{k} = \boldsymbol H^w_{j,k} \hat{\boldsymbol f}'_j. 
\end{equation}
The optimal singular value $\sigma_{1,(j,k)}$ provides a measure of how effectively the forcing $\hat{\boldsymbol f}'_j$ can excite the response at $\hat{\boldsymbol q}'_k$.

\section{Validation of airfoil flow}\label{sec:validation}
In this section, we apply the harmonic resolvent analysis to a flow over a NACA0012 airfoil at an angle of attack of $\alpha = 20^\circ$ and Reynolds number of $\Rey =200$ based on the chord of the airfoil. We set the freestream Mach number at $\Ma_{\infty} = 0.05$, representing incompressible flow regime, to validate our in-house code for compressible flow with the incompressible flow result of \citet{padovan2020analysis} at the same airfoil flow condition. We perform a direct numerical simulation (DNS) to calculate the base flow using a high-fidelity compressible flow solver \textit{CharLES} \citep{bres2017unstructured}, which solves the compressible Navier-Stokes equations using a second-order finite volume method and the third-order Runge--Kutta scheme through an explicit time-stepping method. 

\begin{figure}
  \centerline{\includegraphics{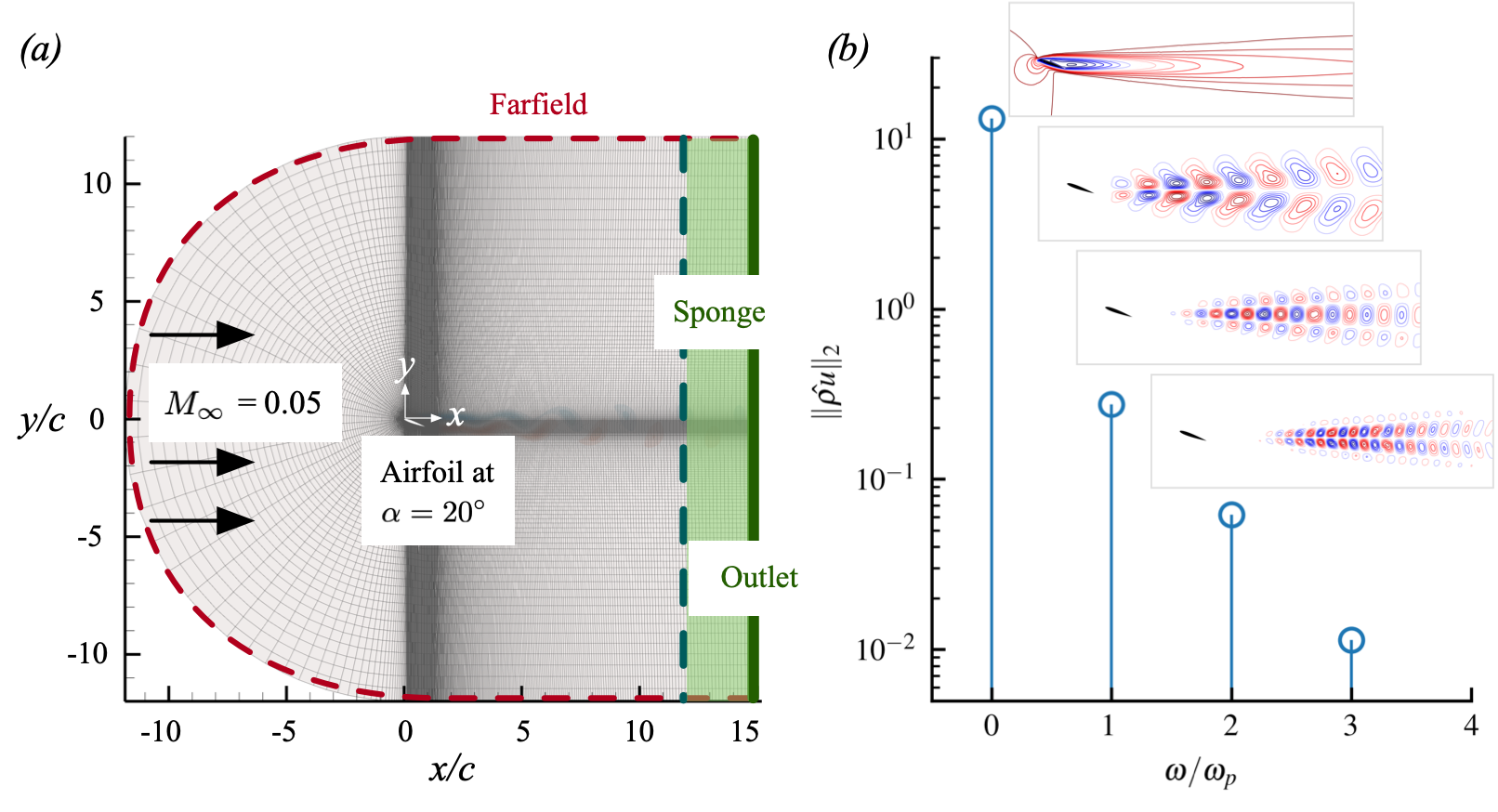}}
   \caption{(a) Computational setup for the DNS of the flow over a NACA0012 airfoil, (b) normalized frequency spectrum of the streamwise momentum and the corresponding Fourier base modes.}
\label{fig:1}
\end{figure}

The computational domain is shown in figure~\ref{fig:1}(a). We take the chord of the airfoil $c$ as a reference length to non-dimensionalize the variables. The origin of the Cartesian coordinate ($x/c=0, y/c=0$) is located at the leading edge of the airfoil. The computational domain has a streamwise extent of $x/c\in[-12,15]$ and the extent in the cross-stream direction is $y/c\in[-12,12]$. We have used a C-shaped mesh with $88,000$ grid points to discretize the computational domain. At the surface of the airfoil, we impose an adiabatic wall boundary condition. We prescribe a characteristic boundary condition with $[\rho,u,v,w,p]=[\rho_{\infty},u_{\infty},0,0,p_{\infty}]$ at the far-field of the domain. Along the domain's outlet, we apply a sponge zone spanning $3c$ in the streamwise direction over the region $x/c\in[12,15]$ to prevent any reflection of the outgoing waves back into the wake of the airfoil. A constant time step $\Delta t U_{\infty}/c=2.0\times 10^{-5}$ is used to advance the simulation in time. 

The simulation ran for a sufficient time so that the flow transients diminished before data collection for analysis. We have gathered data for $45$ convective time ($tU_{\infty}/c$) units and performed the discrete Fourier transform (DFT) to obtain the Fourier coefficients of the base flow states $\hat{\overline{\boldsymbol{q}}}(x,y)$. The base flow is dominated by a periodic vortex shedding in the wake of the airfoil. The norm of the leading streamwise momentum DFT modes is shown in figure~\ref{fig:1}(b). The base flow is periodic with the fundamental frequency ($\omega_p$) corresponding to the vortex shedding mechanism in the wake at the Strouhal number $St=\omega_p c/2\pi u_{\infty}= 0.36$. Also, the presence of energetic components at higher harmonics of the fundamental frequency and the stationary component at the zero frequency is evident in figure~\ref{fig:1}(b).

To perform the harmonic resolvent analysis, we consider a truncated set of base flow frequencies with $\Omega = \{-3,-2,-1,0,1,2,3\}\omega_p$. Consequently, the operator $\hat{\boldsymbol G}$ in equation~(\ref{eqn:G}) will have the Fourier coefficients at the same frequencies present in the base flow set $\Omega$. We generate the discrete operators $\boldsymbol L$ and $\hat{\boldsymbol G}_k$ on a smaller domain with extent $x/c\in[-4,13]$ and $y/c\in[-4,4]$, and approximately $N_g =38,000$ grid points. At the farfield and the surface of the airfoil, we specify the velocity perturbation and the wall-normal gradient of pressure perturbation to zero with an adiabatic condition for the temperature perturbation. We apply a sponge zone over the extent $x/c\in[12,13]$ near the outlet to prevent the outgoing perturbations from reflecting inside the domain. After building the operators $\boldsymbol L$ and $\hat{\boldsymbol G}_k$ we assemble them to form the operator $\boldsymbol T$. The overall size of the operator $\boldsymbol T\in \mathbb{C}^{(5N_g)N_f\times(5N_g)N_f}$ depends on the number of frequencies of the perturbations ($N_f$) and the grid points ($N_g$) used for discretization. Similar to the study of \citet{padovan2020analysis}, we consider the set of frequencies for the perturbation $\tilde{\Omega}=\{-7,\dots,-1,0,1,\dots,7\}\omega_p$ with $N_f= 15$ Fourier coefficients. The assembled operator $\boldsymbol T$ has a size of order $\mathcal{O}(10^6)$ but the structure of $\boldsymbol T$ is highly sparse with the number of non-zero off-diagonal blocks being the same as the number of elements in the set $\Omega$. In contrast, the harmonic resolvent operator $\boldsymbol H$ (i.e., the inverse of $\boldsymbol T$) which has the same size of order $\mathcal{O}(10^6)$ is dense, and its explicit computation is expensive in terms of both CPU hours and storage. Therefore, in practical computation, we use a randomized algorithm \citep{ribeiro2020randomized} to perform the singular value decomposition of the operator $\boldsymbol H$ efficiently. In the randomized algorithm, the action of $\boldsymbol H$ and $\boldsymbol H^*$ can be performed using the sparse operator $\boldsymbol T$ and $\boldsymbol T^*$, respectively, thus, saving the storage and reducing the computation time. We modify the algorithm in the current study to accommodate the additional matrices needed (i.e. matrices $\boldsymbol M$ and $\boldsymbol N$) to perform the SVD of the weighted harmonic resolvent operator $\boldsymbol N \boldsymbol H_p \boldsymbol N^{-1}$. Also, the projection operators discussed at the end of \S\ref{sec:HRA_general} are implemented in the randomized SVD algorithm to remove the phase-shift direction of the operator $\boldsymbol T$ (see Appendix~\ref{appA}). We have used $10$ random test vector to build the low-rank approximation of the operator $T$ \citep{ribeiro2020randomized}. 

\begin{figure}
  \centerline{\includegraphics{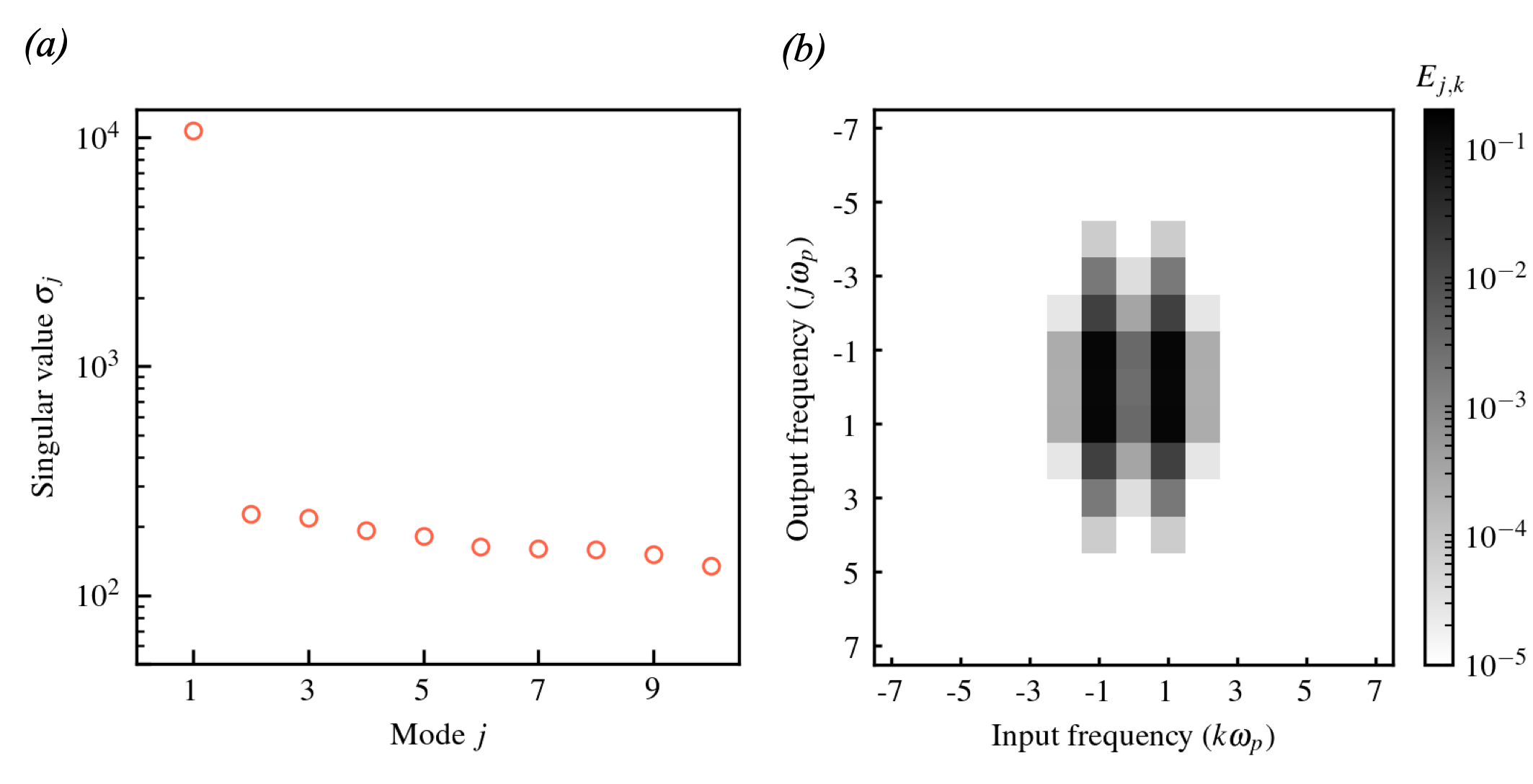}}
   \caption{Singular values of (\textit{a}) the harmonic resolvent operator $\boldsymbol H_p$ and (\textit{b}) fractional variance $E_{j,k}$ corresponding to blocks of $\boldsymbol H_p$.}
\label{fig:2}
\end{figure}

The first $10$ singular values of the weighted harmonic resolvent operator are shown in figure~\ref{fig:2}(a). The order of magnitude significantly drops between the optimal singular value ($\sigma_1$) and the rest, indicating a low-rank behavior of the harmonic resolvent operator, which agrees with the observation in \cite{padovan2020analysis}. Although the exact singular values differ due to the difference in problem setup, the overall trend compares well with the corresponding results of \cite{padovan2020analysis}. The cross-frequency amplification through different blocks of the operator $\boldsymbol H$ is shown in figure~\ref{fig:2}(b). The color of the blocks represents the quantity $E$ defined as follows
\begin{equation}
    E_{j,k} = \frac{\sum_{i=1}^{8} \sigma_{i,(j,k)}^2}{\sum_{i=1}^{8} \sigma_i^2},
    \label{blockE}
\end{equation}
where $\sigma_{i,(k,j)}$ is the $i^{th}$ singular value of the block matrix $\boldsymbol H_{k,j}$, and $\sigma_i$ are the singular values of the complete operator $\boldsymbol H$. The darker color in the plot denotes a significant excitation of the output response at a frequency $j\omega_p$ by the input at frequency $k\omega_p$. In figure~\ref{fig:2}(b), we observe that the flow is responsive to the input at lower frequencies ($<3\omega_p$). In particular, forcing at frequency $\omega_p$ can excite an energetic response in the output at frequencies up to $3\omega_p$. The cross-frequency interaction between perturbations through the base flow manifests in generating the response at frequencies other than the input frequency. The input at higher frequencies ($\geq 3\omega_p$) is ineffective in exciting an energetic response, as evident in figure~\ref{fig:2}(b), indicating a weaker cross-frequency interaction. Such information can not be revealed in the classical resolvent analysis. We note that in the classical resolvent analysis, the time-invariant nature of the base flow does not allow cross-frequency interaction, so the input at a particular frequency will only indicate a response at the same frequency. 

\begin{figure}
   \centerline{\includegraphics{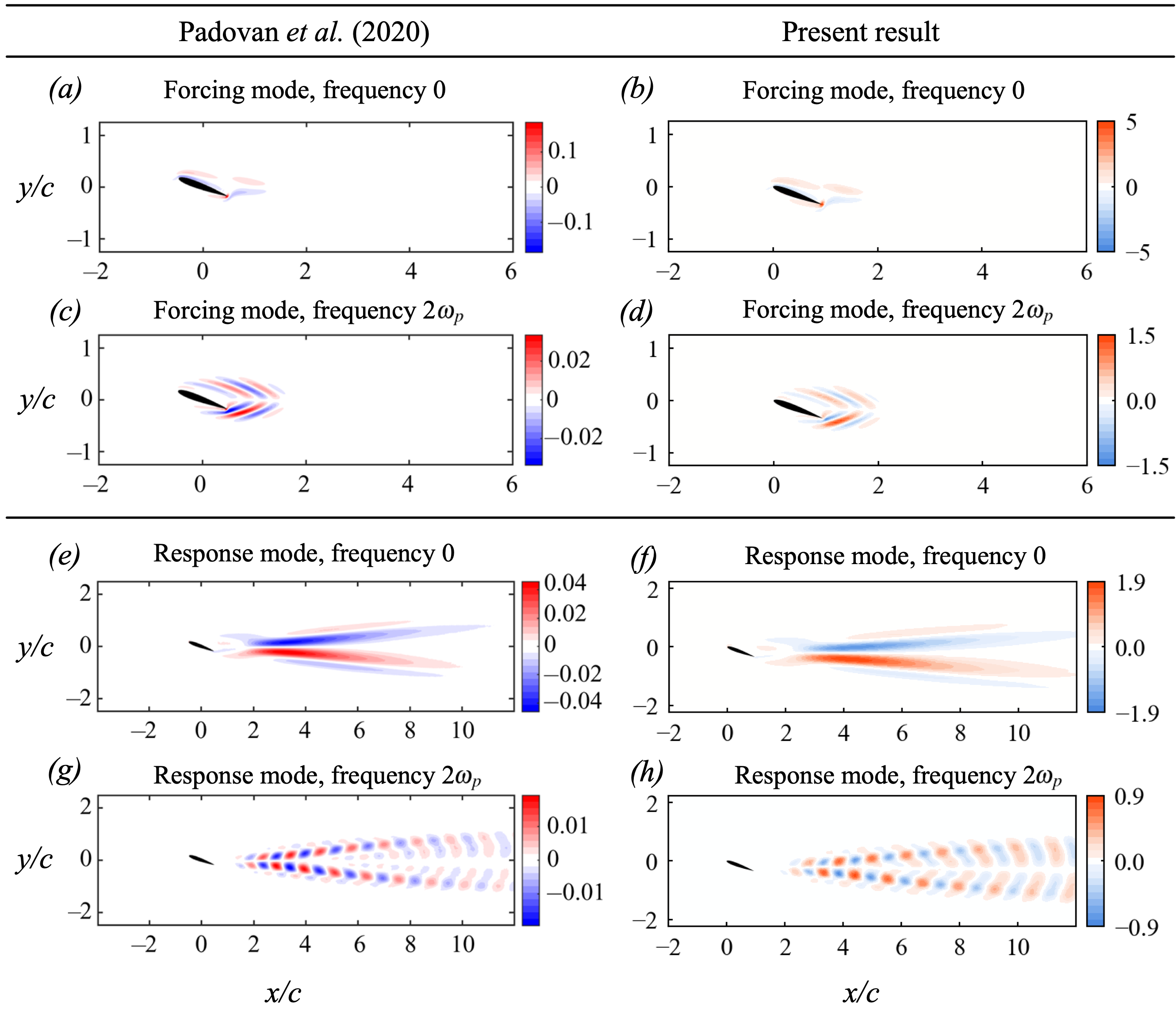}}
  \caption{Comparison of the optimal forcing and response modes (vorticity) at frequencies zero and $2\omega_p$ between the current result and the incompressible flow results by \citet{padovan2020analysis}. }
\label{fig:3}
\end{figure}

The comparison between the optimal forcing and response modes at frequencies $\omega_p$ and $2\omega_p$ reported in \cite{padovan2020analysis}, and the corresponding forcing and response modes obtained in the present work are shown in figure~\ref{fig:3}. The sensitive regions for introducing perturbations are localized around the airfoil as evident in the forcing modes in figure~\ref{fig:3}. The spatial structure of the response modes is located in the wake of the airfoil. Despite the difference in the numerical discretization scheme, the forcing and response mode shapes at both frequencies remarkably agree with the result of \cite{padovan2020analysis}.

\section{Application to flow over open cavity}\label{sec:cavity}
In this section, we consider flow over an open cavity Mach numbers of $\Ma_{\infty}= 0.6$ and $0.8$ to reveal the cross-frequency interactions in high-speed compressible flows. To compute the base flows, we perform DNS of the flow over a rectangular cavity with a length-to-depth ratio of $L/D=2$. Throughout the analysis, we take the depth of the cavity $D$ as a reference length to non-dimensionalize the variables. For both flow configurations, we fix the initial boundary layer momentum thickness ($\theta_0$) at the leading edge of the cavity at $D/\theta_{0}=26.4$. The corresponding Reynolds number for both flows at Mach $0.6$ and $0.8$ based on $\theta_0$ is $Re =56.8$. The computational domain is shown in figure~\ref{fig:4}(a). We place the origin ($x/D=0,y/D=0$) at the leading edge of the cavity. The domain extends $5D$ upstream of the cavity leading edge, and the outlet is placed at a distance of $7D$ from the cavity trailing edge. The domain extends a distance of $9D$ in the wall-normal ($y$) direction. We discretize the computational domain with $1.14\times10^5$ grid points with local mesh refinement near walls and shear layer region. At the surface of the cavity and the upstream and downstream wall, we prescribe adiabatic no-slip boundary conditions. A sponge zone is applied near the outlet and the top boundary with an extent of $1D$ measured from the boundary. At the inlet a characteristics boundary condition with $[\rho,u,v,w,p]=[\rho_{\infty},u_{\infty},0,0,p_{\infty}]$ is specified.

A series of post-transient data is collected over a convective time of $tu_{\infty}/D=75$ to compute the base flow. The unstable shear-layer oscillation due to the Rossiter feedback mechanism \citep{rossiter1964wind,sun2017biglobal} is present in the cavity flow considered. We perform the DFT to get the Fourier coefficients of the base flow states $\hat{\overline{\boldsymbol{q}}}(x,y)$. The normalized spectrum and corresponding DFT modes of the streamwise-momentum component for the cavity flow at Mach $0.6$ reveal the presence of the dominant resonance and its harmonics as shown in figure~\ref{fig:4}(b). The dominant frequency at the Strouhal number of $St=\omega_p L/2\pi u_{\infty}=0.743$ is associated with the Rossiter mode $\text{II}$ based on the semi-empirical formula of oscillatory frequency \citep{rossiter1964wind}.

\subsection{Cavity flow at $M_\infty=0.6$}
We discuss the classical and harmonic resolvent analyses of the cavity flow at Mach 0.6 due to its simplicity of containing only one fundamental frequency. In the classical resolvent analysis, we use the time-averaged mean flow state as the base flow to linearize the governing equations, equivalent to linearizing the equations about the base flow frequency set $\Omega=\{0\}$. In the harmonic resolvent analysis, we consider a set of truncated base flow frequencies of $\Omega=\{-1,0,1\}\omega_p$ leading to the number of base flow frequencies $N_b=3$. We vary the set of perturbation frequencies $\tilde \Omega = \{-m,\dots,-1,0,1\dots,m\}\omega_p$ with $m= 1,3,5$ resulting in the number of perturbation frequencies $N_f= 3,7,$ and $11$, respectively. We build the linear operators for both analyses using a smaller domain with extent $x/D\in[-3,8]$ and $y/D\in [-1,7]$. Approximately $40,000$ grid points are used to discretize the domain. At the cavity surface and the upstream and downstream wall, the velocity perturbations and the wall-normal gradient of the pressure perturbation are specified to be zero. The velocity and density perturbations and gradient of pressure perturbations are specified as zero at the inlet. Sponge layers with an extent of $1D$ are applied near the top and outflow boundaries to dampen the perturbations and prevent any reflection back into the domain. We use $10$ random test vectors to compute the SVD of the harmonic resolvent operator using the randomized algorithm \citep{ribeiro2020randomized}. For the classical resolvent operator, we compute the SVD using the Krylov-based Arnoldi-iteration method with the number of singular values of $4$, Krylov space of $128$ vectors, and a residual tolerance of $10^{-10}$.

\begin{figure}
   \centerline{\includegraphics{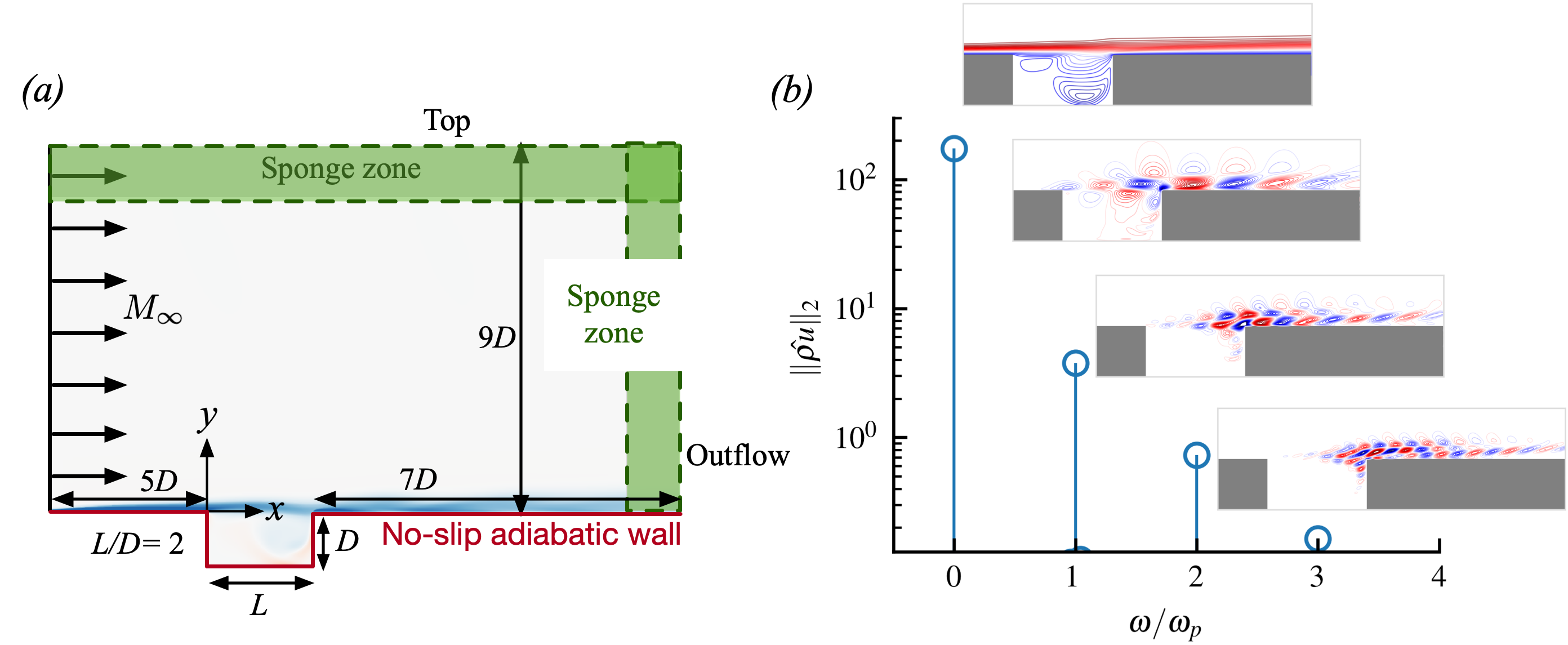}}
  \caption{(a) Computational setup for the DNS of a flow over a rectangular cavity (b) Normalized frequency spectrum of the streamwise momentum and the corresponding Fourier base modes of the cavity flow with $\Ma_{\infty}=0.6$.}
\label{fig:4}
\end{figure}

\begin{figure}
  \centerline{\includegraphics{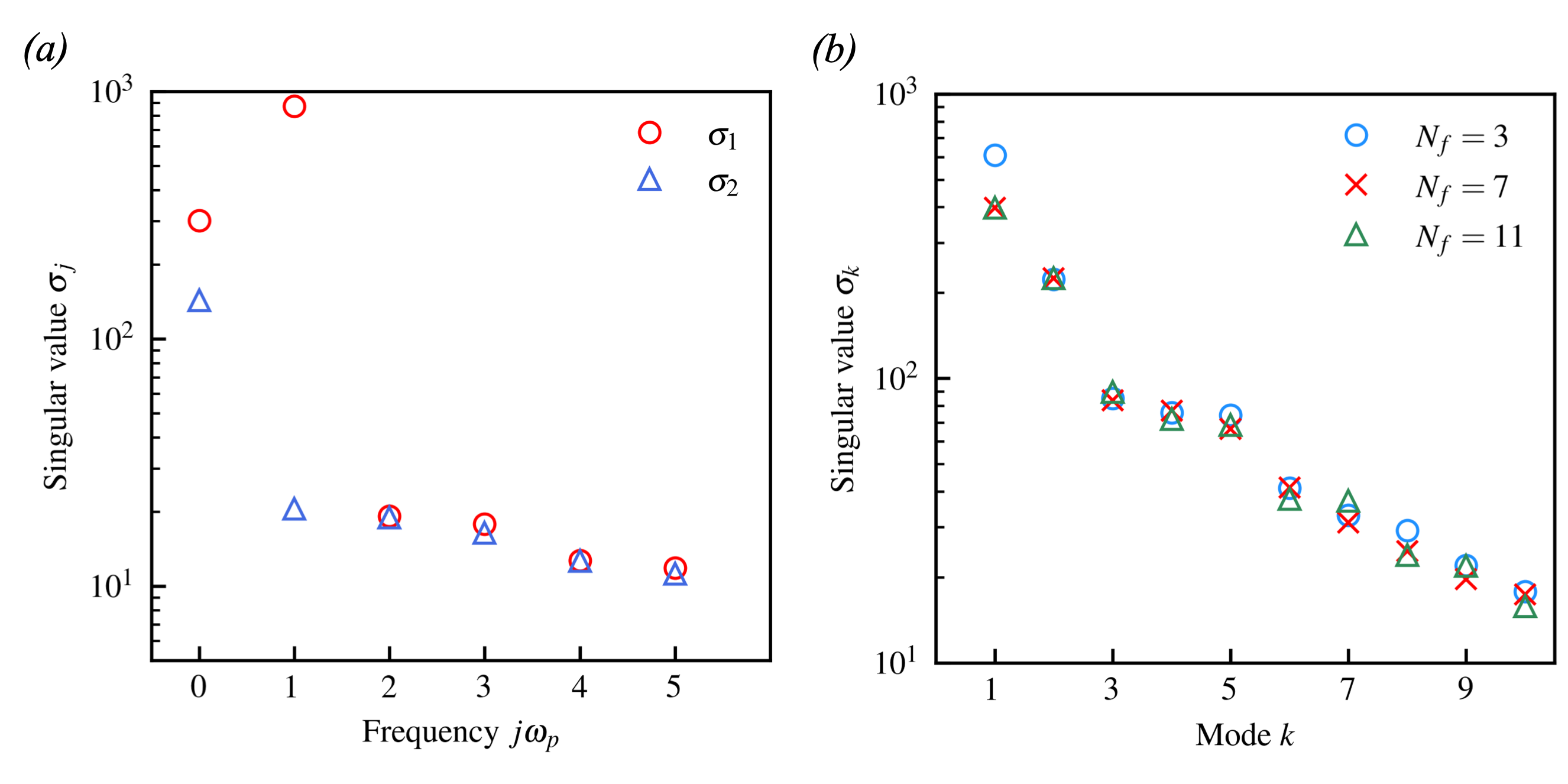}}
  \caption{(a) The first and second singular values of the classical resolvent operator at the fundamental frequency and its harmonics (b) The singular values of the harmonic resolvent operator constructed using $N_b =3$ and $N_f =3, 7, 11$ for the cavity flow at $\Ma_{\infty}= 0.6$.}
\label{fig:5}
\end{figure}

Variation of the first two singular values of the classical resolvent operator as a function of frequency is shown in figure~\ref{fig:5}(a). The flow is most responsive to perturbation at the frequency $\omega_p$ where the large separation between the optimal ($\sigma_1$) and sub-optimal ($\sigma_2$) singular values indicates a rank-1 behavior. Since the leading two singular values overlap at higher frequencies ($\omega\geq \omega_p$), the rank-1 feature is no longer valid. In figure~\ref{fig:5}(b), we plot the first 10 singular values of the harmonic resolvent operator constructed by considering three different numbers of perturbation frequencies ($N_f$) with the same number of base frequencies $N_b=3$. Solving for an increasing number of perturbation frequencies stacked in one singular vector leads to a reduction in the optimal singular value ($\sigma_1$) when $N_f$ increases from 3 to 7, but a further increase of $N_f$ from 7 to 11 implies a minimal effect on capturing the dominant dynamics of the flow as shown in figure~\ref{fig:5}(b). The effect of increasing the perturbation frequencies in the set $\tilde{\Omega}$ on the remaining sub-optimal singular values is trivial. Compared to the singular values of the classical resolvent operator at the dominant frequency ($\omega_p$), the separation between the optimal ($k=1$) and sub-optimal ($k=2$) singular values of the harmonic resolvent operator is smaller. Unlike the classical resolvent operator, the singular values of the harmonic resolvent operator do not reveal the energy amplification of an individual frequency but rather a combined effect of all the coupled frequencies in a set. However, since we solve for cross-frequency modes in one stacked vector, the relative amplitude of each perturbation in the harmonic resolvent mode will reveal their relative dominance, which we will see shortly. 

\begin{figure}
   \centerline{\includegraphics[width=\textwidth]{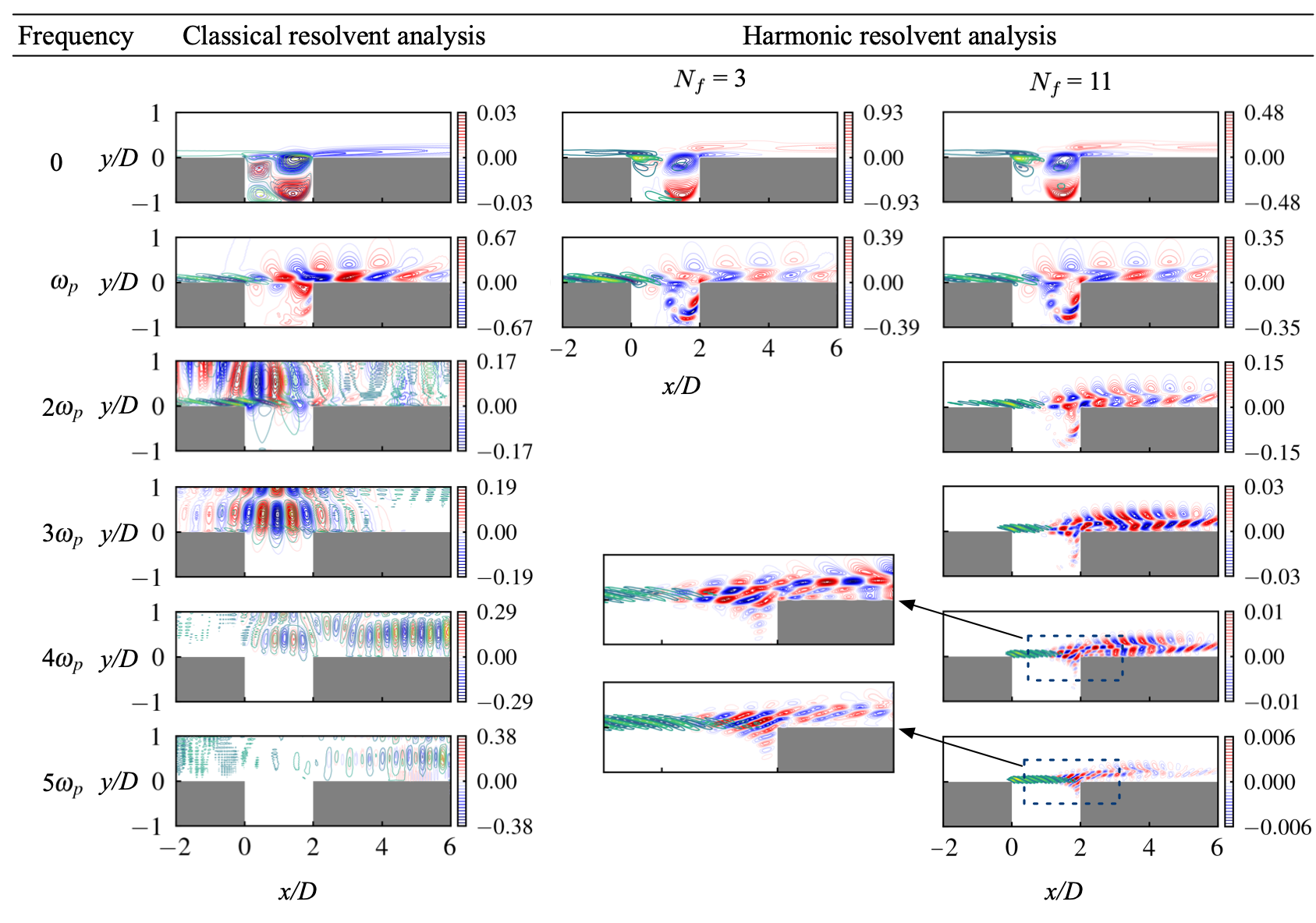}}
  \caption{Real component of the streamwise velocity forcing (blue-green-yellow) and response modes (blue-red) for the cavity flow at $\Ma_{\infty} = 0.6$ associated with $\sigma_1$ at each frequency in figure~\ref{fig:5}(a) for the classical resolvent analysis, and corresponding to $\sigma_1$ in figure~\ref{fig:5}(b) for the harmonic resolvent analysis with $N_b=3$ and $N_f = 3, 11$.}
\label{fig:6}
\end{figure}

To get insight into the coherent structures that get preferentially amplified through linear mechanisms selected by the two transfer functions (i.e., the classical and harmonic resolvent operators), we look into the real component of the streamwise velocity response modes in figure~\ref{fig:6} along with the forcing mode that generates those responses. The forcing and response modes for the classical resolvent analysis correspond to the optimal singular values ($\sigma_1$) at different frequencies. These frequencies, and hence the modes, are decoupled, and each mode oscillates individually in time at the corresponding frequency. Moreover, since the modes are decoupled and follow a unit normalization, the amplitude of each mode can be arbitrary compared to each other. The response mode at frequency zero is mostly located inside the cavity, which resembles the centrifugal mode observed at low frequencies in cavity flow \citep{bres2008three}. The spatial structure of the corresponding forcing mode inside the cavity overlaps with the spatial structures of the response mode indicating the presence of the so-called wavemaker region \citep{symon2018non}. At frequency $\omega_p$, the modal structure represents the unstable Rossiter mode II, with the Kelvin--Helmholtz (KH) instabilities mainly localized in the shear-layer region over the cavity. The forcing mode structures at frequency $\omega_p$ are mostly concentrated near and upstream of the cavity leading edge revealing the convective nature of the KH instabilities. The optimal classical resolvent forcing and response modes at higher frequencies ($\geq 2\omega_p$) are unphysical and provide no meaningful information. The inability to obtain a meaningful mode using the classical resolvent analysis at frequencies where the resolvent operator is not low rank as shown in figure~\ref{fig:5}(a) is a limitation that has been observed in other flows \citep{symon2019tale} before. For an oscillatory flow, which is the case here, \cite{symon2019tale} found the high-rank behavior (i.e., lack of linear amplification mechanism) of the resolvent operator at harmonics of the fundamental frequency of an airfoil flow. After approximating the nonlinear forcing using the triadic interactions of a few highly amplified resolvent response modes, they obtained meaningful structures at those higher harmonics that agreed with the spectral proper orthogonal decomposition modes. In the context of mode's interaction, we speculate that the unphysical classical resolvent modes obtained at the higher harmonics in figure~\ref{fig:6} are due to the absence of modeling cross-frequency interactions among the modes in the classical resolvent formulation. Indeed, we will see next that it is possible to circumvent the limitation very well within the linear framework using the harmonic resolvent analysis without resorting to modeling the nonlinear forcing. 

The streamwise velocity component of the harmonic resolvent forcing and response modes corresponding to the optimal singular value is shown in figure~\ref{fig:6}. The qualitative structure of the forcing and response modes at frequency zero and $\omega_p$ are almost identical regardless of the number of perturbation frequency $N_f$. Since the modes are stacked in one singular vector and orthonormal in their respective inner products, the relative amplitude of the modes at each frequency within one vector varies with the number of perturbation frequencies $N_f$. As the modes are temporally coupled, the mode amplitudes at different frequencies reveal their relative weights of being preferentially excited by an input. For perturbation frequencies with $N_f=3$, the relative weight of the mode at zero frequency is higher than that of the mode at frequency $\omega_p$. With the increase in perturbation frequencies ($N_f=11$), the cross-frequency interaction results in a redistribution of the weights from the zero frequency toward the other frequencies in the set leading to a reduction in the weight at zero frequency. At frequency zero, the organization of the response mode structures is identical to the zero frequency DFT mode inside the cavity shown in figure~\ref{fig:4}(b) with a tail extending over the downstream wall. Unlike the classical resolvent mode, the optimal harmonic resolvent forcing and response mode at frequency zero do not overlap inside the cavity implying a difference in the instability mechanism revealed by both analyses. At frequency $\omega_p$, we observe similar KH mode shapes in the shear-layer region with some qualitative difference near the trailing edge and inside the cavity, compared to the classical resolvent mode, in which the instability also propagates into the cavity following the recirculation contour inside the rear cavity region. Due to the convective nature of the instability at frequency $\omega_p$, the optimal forcing is mostly located upstream of the response mode structures. 

The most significant difference between the classical and harmonic resolvent response modes emerges at the higher harmonics of $\omega_p$ as evident in figure~\ref{fig:6}. The response modes at these higher frequencies resemble the compact KH wavepacket structures with smaller spatial wavelengths along the cavity shear layer, and their concentration shifts toward the trailing edge with increasing frequency. Meanwhile, the relative amplitude of each mode decreases as frequency increases. Since these instability modes at high frequencies ($\geq 2\omega_p$) are also convective, the forcing structures are mainly concentrated upstream of the response. With an increase in frequency, the most responsive region to introduce forcing shifts towards the shear-layer region after the cavity leading edge. Generation of harmonics is a nonlinear process, and by incorporating the base flow at frequency $\omega_p$ into the set about which the linearization is done, we successfully recover physical modes at those higher frequencies from the harmonic resolvent analysis. Again we stress that the cross-frequency interaction between perturbations at different frequencies through the base flow is the underlying mechanism for this outcome.

\begin{figure}
  \centerline{\includegraphics{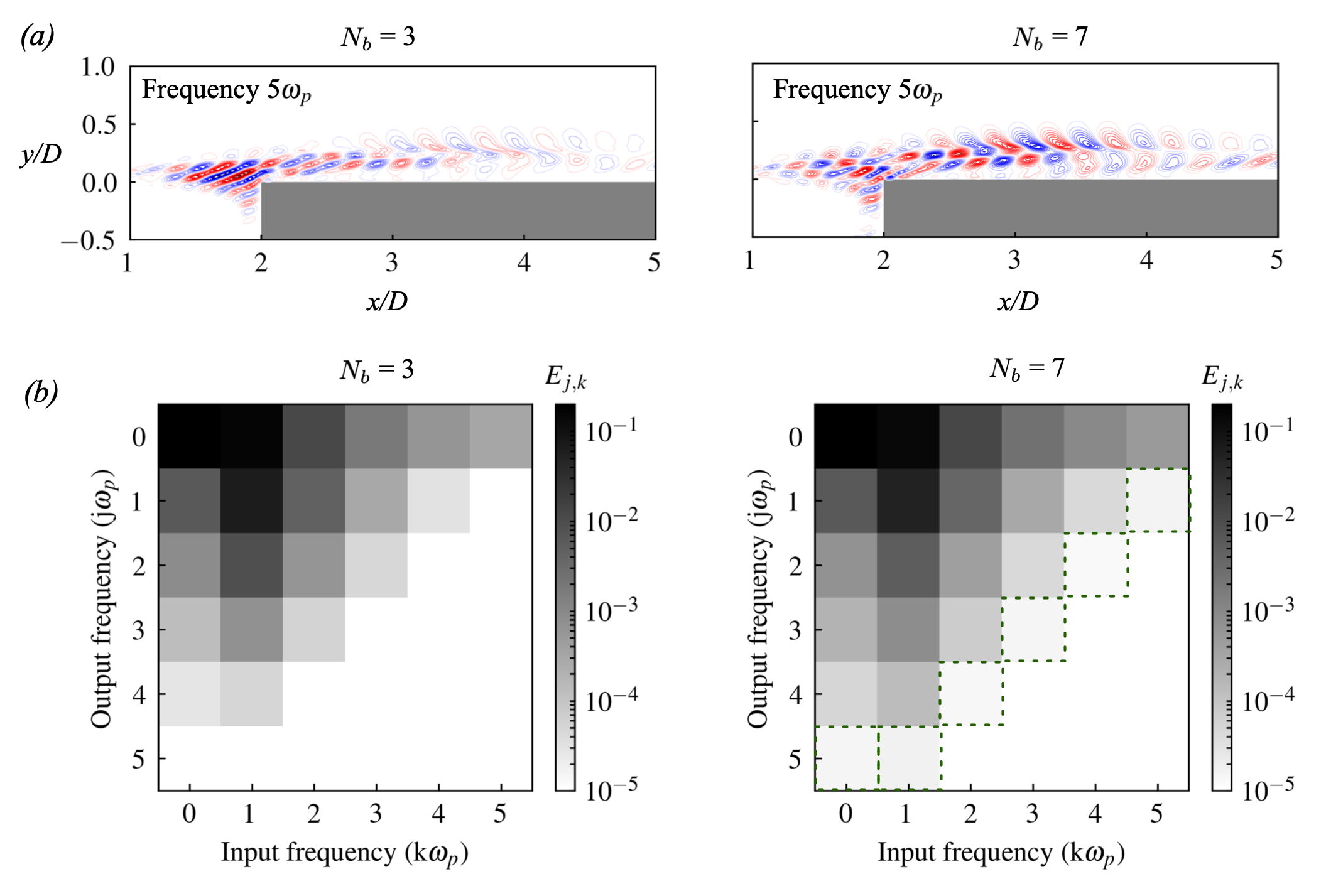}}
  \caption{(a) Real component of the optimal streamwise velocity response modes for the cavity flow ($\Ma_{\infty} = 0.6$) at frequency $5\omega_p$  and (b) fractional variance $E_{j,k}$ of block singular values of the harmonic resolvent operator obtained using $N_b= 3, 7$ and $N_f= 11$.}
\label{fig:7}
\end{figure}

To understand the effect of the base flow frequency truncation on the cross-frequency interaction, we perform the harmonic resolvent analysis by increasing the number of base frequencies in the set $\Omega$ from $N_b=3$ to $N_b=7$, i.e., considering the base flow frequency set $\Omega = \{-3,-2,-1,0,1,2,3\}\omega_p$. We keep the number of perturbation frequencies to $N_f=11$ as a constant. The resulting singular value distribution of the harmonic resolvent operator in both cases is found to be not significantly different. The structures of the response modes at frequencies between zero and $3\omega_p$ have almost identical shapes with minor variations in the amplitude, which we do not show here for brevity. However, the increase of base flow frequencies in the modeling affects the modal structures at frequencies above $3\omega_p$. In particular, a closer inspection of the modal structures near the trailing edge of the cavity in figure~\ref{fig:7}(a) for $N_b=7$ shows an increase in small-scale structures at frequency $5\omega_p$ compared to the corresponding modes obtained using $N_b=3$. The increase in the base frequencies in the set $\Omega$ extends the cross-frequency coupling between the perturbations through $\hat{\boldsymbol G}_k$ according to equation~(\ref{freq_coupling}). Consequently, it adds more paths to the cascaded energy transfer process from the low frequency towards the higher frequency modes and, hence a more accurate representation of the modal structures at those frequencies (i.e., $4\omega_p$ and $5\omega_p$) can be obtained. 

To examine the input-output amplifications from different frequency pairs, we obtain the cross-frequency interaction in the set $\tilde \Omega$ through the block singular values of the harmonic resolvent operator (see \S\ref{sec:modal}). We reconstruct a low-rank approximation of the harmonic resolvent operator using the singular values (and the corresponding singular vectors) from $k=1$ to $k=8$. Then by performing the SVD of individual blocks of the harmonic resolvent operator, we compute the quantity $E_{j,k}$ following equation~(\ref{blockE}). The result is shown in figure~\ref{fig:7}(b), where the darker color represents the significant coupling between the pair of frequencies ($j\omega_p,k\omega_p$). Looking into each row of the input-output map, we notice the additional presence of blocks implying an increase in the extent of the cross-frequency coupling. Comparing the energy in these additional non-zero blocks (highlighted by green dashed lines) for rows corresponding to frequencies less than $4\omega_p$, we expect a negligible effect on modeling the dominant dynamics at those frequencies by increasing $N_b$ from $3$ to $7$, because their amplitudes are insignificant compared to other blocks in the same row. However, the energy in the additional interaction blocks can be significant for higher frequencies ($\geq 4\omega_p$) since those are the dominant interactions affecting those frequencies, as evident in row $5$ of the map for $N_b=7$.

The analysis that was performed until now reveals the effect of truncation of both base flow and the perturbation frequencies on capturing the dominant dynamics of perturbations. The truncation of perturbation frequencies ($N_f$) affects the relative amplitude of the modes to some extent but minimally affects the spatial structures of the modes. The truncation of the frequencies in the base flow set about which linearization is performed influences the modal structures at relatively high frequencies. Consequently, the choice of truncation of base flow frequency depends on the analysis objective. If one wishes to model the structures accurately at high frequencies, including more frequencies in the base flow set will enhance the accuracy. Otherwise, if the objective is to get insight into the dominant amplified structures in the flow linearizing about a few energetic frequencies should suffice.

\subsection{Cavity flow at $M_\infty=0.8$}
\begin{figure}
   \centerline{\includegraphics{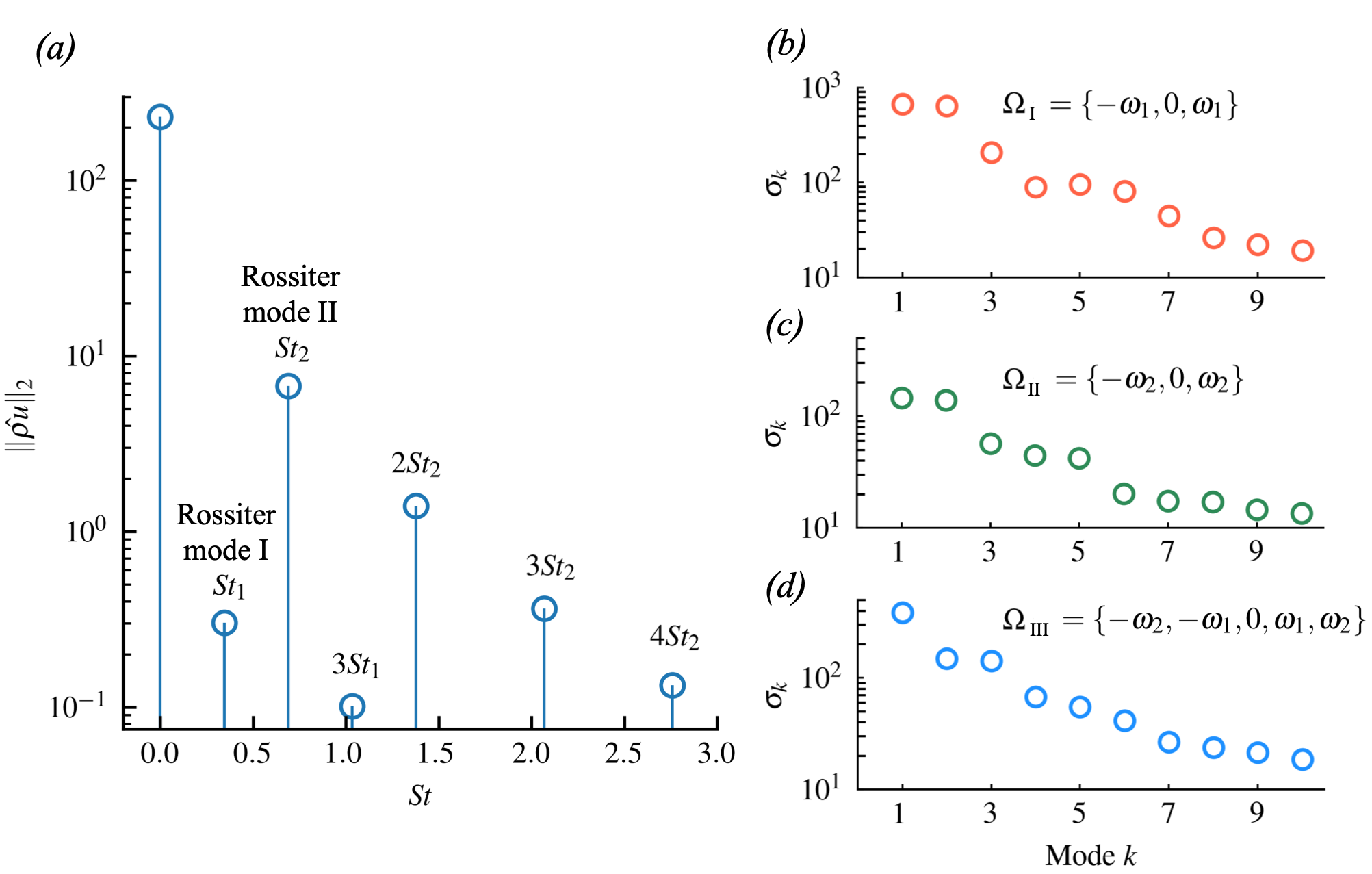}}
  \caption{(a) Frequency spectrum of the streamwise momentum of cavity flow at Mach 0.8. (b-d) The leading $10$ singular values of the harmonic resolvent operator constructed using the sets of base flow frequencies $\Omega_{\text{I}}=\{-\omega_1,0,\omega_1\}$, $\Omega_{\text{II}}=\{-\omega_2,0,\omega_2\}$, $\Omega_{\text{III}}=\{-\omega_2,-\omega_1,0,\omega_1,\omega_2\}$. }
\label{fig:8}
\end{figure}

We also apply the harmonic resolvent analysis to understand the cross-frequency interaction in the cavity flow at Mach $0.8$ which contains more than one resonance. We collect post-transient data and perform DFT to obtain the base flow $\hat{\overline{\boldsymbol{q}}}(x,y)$ as before. The frequency spectrum of the streamwise momentum component of the base flow is shown in figure~\ref{fig:8}(a). Two coexisting Rossiter mechanisms drive the base flow oscillation with the frequency of the Rossiter mode $\text{I}$ is $St_1=\omega_1 L/2\pi u_{\infty}=0.345$ and the frequency of the Rossiter mode $\text{II}$ corresponds to $St_2=\omega_2 L/2\pi u_{\infty}=0.689$, which agrees well with oscillatory frequency based on the semi-empirical formula of Rossiter \citep{rossiter1964wind}. In addition, we observe the presence of the harmonics of $St_1$ and $St_2$ in the spectrum. Unlike the cavity flow at Mach 0.6, the spectrum of the base flow at Mach 0.8 is not monochromatic and thus poses several ways to construct the set of the base flow frequency $\Omega$. Here we consider three sets of base flow frequency $\Omega_{\text{I}} = \{-\omega_1,0,\omega_1\}$, $\Omega_{\text{II}} = \{-\omega_2,0,\omega_2\}$, and $\Omega_{\text{III}}=\{-\omega_2,-\omega_1,0,\omega_1,\omega_2\}$ to approximate the time-varying base flow and linearize the dynamics about those frequency sets. We remark here that in this particular flow, the frequency of Rossiter mode $\text{II}$ ($\omega_2$) is approximately two times the frequency of Rossiter mode $\text{I}$ ($\omega_1$). Thus we can combine them in the set $\Omega_{\text{III}}$ without modifying the theoretical formulation. To perform the harmonic resolvent analysis, we choose the set of perturbation frequencies to be $\tilde \Omega = \{-5,\dots,\-1,0,1,\dots,5\}\omega_{p}$, where $\omega_p = \omega_1$ for the base flow frequency sets $\Omega_{\text{I}}$ and $\Omega_{\text{III}}$, and $\omega_p = \omega_2$ for the set $\Omega_{\text{II}}$. We follow the same steps used for the cavity flow at Mach $0.6$ to construct the linear operators $\boldsymbol L$ and $\hat{\boldsymbol G}_k$ and to perform the SVD of the harmonic resolvent operator. 

\begin{figure}
   \centerline{\includegraphics{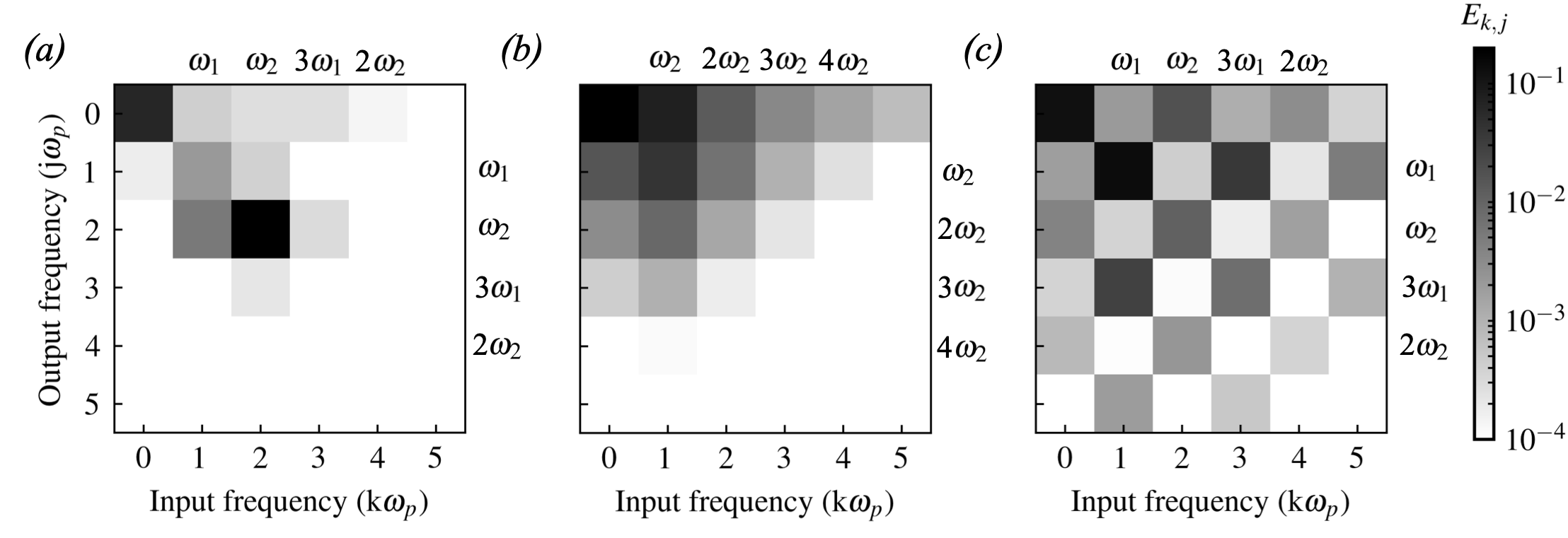}}
  \caption{Fractional variance $E_{j,k}$ of block singular values of the harmonic resolvent operator constructed using the sets of base flow frequencies (a) $\Omega_{\text{I}}=\{-\omega_1,0,\omega_1\}$, (b) $\Omega_{\text{II}}=\{-\omega_2,0,\omega_2\}$, and (c) $\Omega_{\text{III}}=\{-\omega_2,-\omega_1,0,\omega_1,\omega_2\}$.}
\label{fig:9}
\end{figure}

The first $10$ of singular values of the harmonic resolvent operator obtained by linearizing about the three base flow frequency sets are shown in figure~\ref{fig:8}(b-d). The optimal singular values ($\sigma_1$) in all three cases are greater than the singular value $\sigma_{10}$ by approximately an order of magnitude. To analyze the cross-frequency interactions we reconstruct the harmonic resolvent operator using singular values (and associated singular vectors) from $k= 1$ to $k= 8$ in figure~\ref{fig:8}(b-d). The fractional variance of the block singular values $E_{j,k}$ of the harmonic resolvent operators calculated according to equation~(\ref{blockE}) is plotted in figure~\ref{fig:9}. In figure~\ref{fig:9}(a) the map shows that the self-interactions at frequencies zero, and $\omega_2$ dominate the flow by amplification through the diagonal blocks whereas the cross-frequency interactions are weak. Linearizing about the base flow frequency set $\Omega_{\text{I}}$ does not model the cross-frequency interactions effectively as the Rossiter mode $\text{I}$ is not dominant in the nonlinear flow. When we linearize about the set $\Omega_{\text{II}}$ to perform the harmonic resolvent analysis the frequency interaction map exhibits a progressive pattern with an increase in off-diagonal blocks in figure~\ref{fig:9}(b). The map shows the amplification of perturbations at harmonics of Rossiter mode $\text{II}$ through the cascaded cross-frequency interactions. The base flow frequency set $\Omega_{\text{III}}$ contains both Rossiter mechanisms present in the nonlinear flow, and the frequency interaction map is shown in figure~\ref{fig:9}(c). The input-output pair of frequencies ($\omega_1, \omega_1$), ($\omega_1,3\omega_1$), and ($\omega_1,5\omega_1$) are strongly coupled such that the interactions affect the perturbation amplification at odd harmonics of the Rossiter mode $\text{I}$. Whereas the interactions between the frequency pairs ($\omega_1, \omega_2$) and ($\omega_1,2\omega_2$) are negligible indicating that perturbation at Rossiter mode $\text{I}$ frequency does not interact significantly with perturbation at Rossiter mode $\text{II}$ frequency. A similar conclusion can be drawn by observing the frequency $\omega_2$ and $2\omega_2$ rows of the map. The perturbations at the harmonics of both Rossiter modes are amplified through the interactions among corresponding Rossiter modes and the mean. Alternatively, based on the feature of the column of frequency $\omega_1$, we can also find that the input perturbation at Rossiter mode $\text{I}$ frequency $\omega_1$ will generate a significant response at frequencies $\omega_1$, $3\omega_1$, and $5\omega_1$ but no response at the Rossiter mode $\text{II}$ frequency $\omega_2$, and vice-versa.

\section{Conclusion}\label{sec:conclusion}
This paper presents the harmonic resolvent analysis framework for compressible flows. We linearized the compressible Navier-stokes equation around a time-varying base flow and used Fourier expansions to obtain the frequency domain input-output relation of the perturbations in the compressible flow governed by the harmonic resolvent operator. Due to the higher-order nonlinearity in the compressible NSE compared to the incompressible form, the linearized equation contains nonlinear products among the base state variables. As a result, the product between base variables in the time domain leads to the evaluation of multiple convolutions in the frequency domain representation adding complexity to the modeling framework of the compressible version compared to its incompressible counterpart. We discussed the Fourier expansions of different terms of the linearized equation and detailed the process of constructing the harmonic resolvent operator.

The SVD of the harmonic resolvent operator provides a way to measure the amplification of perturbations in the flow considering frequency interactions. Both classical and harmonic resolvent analysis identify the instability mechanisms at the dominant frequency observed in the nonlinear flow as the convective Kelvin--Helmholtz nature. However, the absence of cross-frequency interaction modeling in the classical resolvent formulation leads to the identification of spurious modal structures at the harmonics of the dominant frequency. The limitation was overcome in the harmonic resolvent analysis and the physically meaningful structures at those frequencies are recovered. The SVD of the individual block matrices within the harmonic resolvent operator provides additional information about the extent of the cross-frequency interactions among the perturbation through the base flow. Using the singular values of each block we generated the cross-frequency interaction map which revealed the cascaded path of frequency interaction and energy transfer from the fundamental toward the higher harmonics for cavity flow with one resonance mechanism. Utilizing the same method allowed us to investigate the nature of cross-frequency interaction in cavity flow with two different resonant mechanisms. Although we used both resonant frequencies simultaneously to model the base flow, the harmonic resolvent analysis correctly identified the frequencies as two distinct physical mechanisms and not the harmonics of one another. 

\backsection[Acknowledgements]{We would like to thank Dr.~Alberto Padovan for the insightful discussion on the computational steps of the SVD of the harmonic resolvent operator. We acknowledge the computing facility provided by Syracuse University Research Computing. We also gratefully acknowledge the support provided by Syracuse University.}

\backsection[Funding]{This material is based upon work supported by the Air Force Office of Scientific Research under award number FA9550-24-1-0136 (Program Officer: Dr.~Gregg Abate).}

\backsection[Declaration of interests]{The authors report no conflict of interest.}

\appendix

\section{Singularity of the operator T}\label{appA}
If the base state $\boldsymbol Q(t)$ is an exact solution of the equation~(\ref{NDS}) then $\boldsymbol f'(t)$ becomes zero in equation~(\ref{LDS}). Since $\boldsymbol Q(t+\tau)$ is also an exact solution for any particular time shift $\tau$, the time derivative of the base state $d\boldsymbol Q/dt$ satisfies the unforced dynamics in equation~(\ref{LDS}). We can verify the statement by expanding $\boldsymbol Q(t+\tau)$ using the Taylor series approximation for a small shift $\tau$ as
\begin{equation}
    \boldsymbol Q(t+\tau) = \boldsymbol Q(t) + \frac{d\boldsymbol Q(t)}{dt} \tau,
\end{equation}
and substituting the expression in the equation~(\ref{NDS})
\begin{equation}
    \frac{d\boldsymbol Q(t)}{dt} + \tau \frac{d}{dt} (\frac{d \boldsymbol Q(t)}{dt}) = \mathcal N(\boldsymbol Q(t)) + \left. \frac{\partial \mathcal N}{\partial \boldsymbol q} \right\vert_{\boldsymbol Q(t)} \frac{d \boldsymbol Q(t)}{dt} \tau, 
\end{equation}
which after rearranging gives
\begin{equation}
    \frac{d}{dt} \left(\frac{d \boldsymbol Q(t)}{dt}\right) = \boldsymbol A(t) \frac{d \boldsymbol Q(t)}{dt}.
    \label{unforced_LDS}
\end{equation}
By expanding $d\boldsymbol Q(t)/dt$ and $\boldsymbol A(t)$ in their respective Fourier series and substituting in equation~(\ref{unforced_LDS}) we get
\begin{equation}
    \boldsymbol T\, \widehat{\frac{d\boldsymbol Q}{dt}} = 0,
\end{equation}
where $\widehat{d\boldsymbol Q/dt}$ is the vector containing the Fourier coefficients of $d\boldsymbol Q/dt$. Thus the vector $\widehat{d\boldsymbol Q/dt}$ is in the right null space of the operator $\boldsymbol T$, which makes $\boldsymbol T$ singular and non-invertible.

\section{Frequency domain representation of the linearized flux terms}\label{appB}
We have split the terms of the linearized Euler flux vector into two parts $\mathcal G_j^e(\overline{\boldsymbol q}(t),\boldsymbol q'(t))$ and $\boldsymbol L \boldsymbol q'(t)$ in equation~(\ref{LNSE2}) to make the transformation from the time domain to the frequency domain more tractable. The terms in $\mathcal G_j^e(\overline{\boldsymbol q}(t),\boldsymbol q'(t))$ read
\begin{equation}
\setlength{\arraycolsep}{0pt}
\renewcommand{\arraystretch}{2.5}
\mathcal G_j^e = \left[
\begin{array}{c}
    0\\
   \displaystyle
   \frac{\overline m_i}{\overline \rho} m'_j + \frac{\overline m_j}{\overline \rho} m'_i - \frac{\overline m_i \overline m_j}{\overline{\rho}^2} \rho' + \frac{\alpha}{2} \frac{\overline m_k \overline m_k}{ \overline{\rho}^2} \rho' \delta_{ij} - \alpha \frac{\overline m_k}{\overline \rho} m_k'\delta_{ij}\\
   \displaystyle
    \gamma \frac{\overline{\rho E}}{\overline \rho} m_j' - \frac{\alpha}{2} \frac{\overline m_k \overline m_k}{\overline \rho^2} m_j' - \gamma \frac{\overline{\rho E}\, \overline m_j}{\overline \rho^2} \rho' + \alpha \frac{\overline m_k \overline m_k \overline m_j}{\overline \rho^3} \rho' + \gamma \frac{\overline m_j}{\overline \rho} (\rho E)' - \alpha \frac{\overline m_j \overline m_k}{\overline \rho^2} m_k'
\end{array}  \right] ,
\label{B1}
\end{equation}
where $\alpha = \gamma-1$. The operator $\boldsymbol L$ is given by
\begin{equation}
\setlength{\arraycolsep}{10pt}
\renewcommand{\arraystretch}{1.5}
\boldsymbol L = \frac{\partial}{\partial x_j}\left[
\begin{array}{ccc}
    0& \delta_{ij}& 0\\
   \displaystyle
   0& 0& (\gamma-1)\delta_{ij}\\
   \displaystyle
    0& 0& 0
\end{array}  \right].
\end{equation}
Now let us show the Fourier expansion of the linearized Euler flux terms in $\mathcal G_j^e(\overline{\boldsymbol q}(t),\boldsymbol q'(t))$. We will detail the conversion of the terms in the second and third row of equation~(\ref{B1}) from the time domain to the frequency domain. Substituting the Fourier series expansion of equation~(\ref{FS}) in equation~(\ref{B1}) we get
\begin{align}
    &\sum_{\substack{l\in\tilde \Omega\\(p-l)\in \Omega}} \widehat{\left(\frac{\overline m_i}{\overline \rho}\right)}_{p-l} m'_{j,l} + \widehat{\left(\frac{\overline m_j}{\overline \rho}\right)}_{p-l} m'_{i,l} - \widehat{\left(\frac{\overline m_i \overline m_j}{\overline{\rho}^2}\right)}_{p-l} \rho'_l + \frac{\alpha}{2} \widehat{\left(\frac{\overline m_k \overline m_k}{ \overline{\rho}^2}\right)}_{p-l} \rho'_l \delta_{ij} \nonumber\\
    &- \alpha \widehat{\left(\frac{\overline m_k}{\overline \rho}\right)}_{p-l} m'_{k,l}\delta_{ij}\\[3pt]
     &\sum_{\substack{l\in\tilde \Omega\\(p-l)\in \Omega}} \gamma\widehat{\left(\frac{\overline{\rho E}}{\overline \rho}\right)}_{p-l} m_{j,l}' - \frac{\alpha}{2} \widehat{\left(\frac{\overline m_k \overline m_k}{\overline \rho^2}\right)}_{p-l} m_{j,l}' - \gamma \widehat{\left(\frac{\overline{\rho E}\, \overline m_j}{\overline \rho^2}\right)}_{p-l} \rho'_l + \alpha \widehat{\left(\frac{\overline m_k \overline m_k \overline m_j}{\overline \rho^3}\right)}_{p-l} \rho'_l \nonumber \\
    &+ \gamma \widehat{\left(\frac{\overline m_j}{\overline \rho}\right)}_{p-l} (\rho E)'_l - \alpha \widehat{\left(\frac{\overline m_j \overline m_k}{\overline \rho^2}\right)}_{p-l} m_{k,l}'
\end{align}

with the frequency of the base flow terms being calculated as
\begin{align}
    \widehat{\left(\frac{\overline m_i}{\overline \rho}\right)}_{p-l} &= \sum_{a \in \Omega}  \hat{\overline{r}}_{p-l-a}\, \hat{\overline{m}}_{i,a},\\[3pt]
    \widehat{\left(\frac{\overline m_i \overline m_j}{\overline{\rho}^2}\right)}_{p-l} &= \sum_{a \in \Omega} \sum_{o \in \Omega} \sum_{q \in \Omega}  \hat{\overline{r}}_{p-l-a-o-q}\, \hat{\overline{r}}_{a}\, \hat{\overline{m}}_{i,o}\, \hat{\overline{m}}_{j,q},\\[3pt]
    \widehat{\left(\frac{\overline {\rho E}}{\overline \rho}\right)}_{p-l} &= \sum_{a \in \Omega}  \hat{\overline{r}}_{p-l-a}\, \hat{\overline{\rho E}}_{a},\\[3pt]
    \widehat{\left(\frac{\overline{\rho E} \overline m_j}{\overline{\rho}^2}\right)}_{p-l} &= \sum_{a \in \Omega} \sum_{o \in \Omega} \sum_{q \in \Omega}  \hat{\overline{r}}_{p-l-a-o-q}\, \hat{\overline{r}}_{a}\, \hat{\overline{\rho E}}_{o}\, \hat{\overline{m}}_{j,q},\\[3pt]
    \widehat{\left(\frac{\overline m_k \overline m_k \overline m_j}{\overline{\rho}^3}\right)}_{p-l} &= \sum_{a \in \Omega} \sum_{b \in \Omega} \sum_{o \in \Omega} \sum_{s \in \Omega} \sum_{q \in \Omega}  \hat{\overline{r}}_{p-l-a-b-o-s-q}\, \hat{\overline{r}}_{a}\, \hat{\overline{r}}_{b}\, \hat{\overline{m}}_{k,o}\, \hat{\overline{m}}_{k,s}\, \hat{\overline{m}}_{j,q},
\end{align}
where we define $\hat{\overline r}$ as the Fourier coefficients of the term $1/\overline{\rho}(t)$. Following the same procedure we perform Fourier expansion of the linearized viscous flux terms in equation~(\ref{coupledeq2}).

\section{Details of the primitive to conservative variable transformation matrix}\label{appC}
The primitive perturbation variables $\boldsymbol q_p'(t) = [\rho',u'_i,T']$ can be transformed into the  conservative perturbation variables $\boldsymbol q'(t) = [\rho',m'_i,(\rho E)']$ as follows
\begin{equation}
\setlength{\arraycolsep}{8pt}
\renewcommand{\arraystretch}{3.0}
\left[
\begin{array}{c}
    \rho'\\
   \displaystyle
   m'_i\\
   \displaystyle
   (\rho E)'
\end{array}  \right] =
\underbrace{\left[
\begin{array}{ccc}
    1& 0& 0\\
   \displaystyle
   \frac{\overline{m}_i}{\overline{\rho}}& \overline{\rho}& 0\\
   \displaystyle
   \frac{\overline{\rho E}}{\overline{\rho}}& \overline{m}_i& \frac{\overline{\rho}}{\gamma (\gamma-1) \Ma^2}
\end{array}  \right]}_{\boldsymbol S} 
\left[
\begin{array}{c}
    \rho'\\
   \displaystyle
   u'_i\\
   \displaystyle
   T'
\end{array}  \right].
\end{equation}
In the frequency domain, the transformation can be represented as 
\begin{equation}
\setlength{\arraycolsep}{5pt}
\renewcommand{\arraystretch}{1.6}
\left[
\begin{array}{c}
    \vdots\\
    \hat{\boldsymbol q'}_{-1}\\
   \displaystyle
   \hat{\boldsymbol q'}_0\\
   \displaystyle
   \hat{\boldsymbol q'}_1\\
   \vdots
\end{array}  \right] = 
\underbrace{\left[
\begin{array}{ccccc}
 \ddots &\vdots& \vdots& \vdots&   \\
  \displaystyle
  \dots& \hat{\boldsymbol S}_0 & \hat{\boldsymbol S}_{-1}& \hat{\boldsymbol S}_{-2} & \dots \\
  \displaystyle
  \dots& \boldsymbol S_1 & \boldsymbol S_0& \boldsymbol S_{-1} &\dots  \\
  \displaystyle
  \dots& \boldsymbol S_2& \boldsymbol S_1 & \boldsymbol S_{0} &\dots\\
  \displaystyle
   &\vdots& \vdots& \vdots& \ddots
\end{array}  \right]}_{M} 
\left[
\begin{array}{c}
    \vdots\\
    \hat{\boldsymbol q'}_{p,-1}\\
   \displaystyle
   \hat{\boldsymbol q'}_{p,0}\\
   \displaystyle
   \hat{\boldsymbol q'}_{p,1}\\
   \vdots
\end{array}  \right],
\end{equation}
where $\hat{\boldsymbol S}_k$ denotes the Fourier coefficients of the matrix $\boldsymbol S$. The number of nonzero off-diagonal blocks in $\boldsymbol M$ corresponds to the number of base flow frequencies in the set $\Omega$.

\bibliographystyle{unsrtnat}
\bibliography{references}  






\end{document}